\documentclass[proc]{edpsmath}
\newtheorem{lemma}{Lemma}[section]

\newtheorem{remark}[lemma]{Remark}

\usepackage{graphicx}
\usepackage{url}
\usepackage{color}
\definecolor{marin}{rgb}   {0.,   0.3,   0.7} 
\definecolor{rouge}{rgb}   {0.8,   0.,   0.} 
\definecolor{sepia}{rgb}   {0.8,   0.5,   0.} 
\usepackage[colorlinks,citecolor=marin,linkcolor=rouge,
            bookmarksopen,
            bookmarksnumbered
           ]{hyperref}

\usepackage{float}
\usepackage{dsfont}

\begin{document}

\title{Vlasov on GPU (VOG project)}\thanks{Thanks to Edwin Chacon-Golcher, Philippe Helluy, Guillaume Latu, Pierre Navaro for fruitful discussions and helps}
\thanks{Thanks to the CEMRACS organizers and participants for the nice stay}
\thanks{This work was carried out within the framework the European Fusion Development Agreement
and the French Research Federation for Fusion Studies. It is supported by the European Communities
under the contract of Association between Euratom and CEA. The views and opinions expressed
herein do not necessarily reflect those of the European Commission.}
%
\author{M. Mehrenberger}\address{IRMA, Universit\'e de Strasbourg, 7, rue Ren\'e Descartes, F-67084 Strasbourg \& INRIA-Nancy Grand-Est, projet CALVI,
\email{mehrenbe@math.unistra.fr}.}
\author{C. Steiner}\address{IRMA, Universit\'e de Strasbourg, 7, rue Ren\'e Descartes, F-67084 Strasbourg \& INRIA-Nancy Grand-Est, projet CALVI,
\email{steiner@math.unistra.fr}.}
\author{L. Marradi}\address{LIPHY, Universit\'e Joseph Fourier, 140, avenue de la Physique, F-38402 Saint Martin d'H\`eres,
\email{ luca.marradi@ujf-grenoble.fr}.}
\author{N. Crouseilles}\address{INRIA-Rennes Bretagne Atlantique, projet IPSO \& IRMAR, Universit\'e de Rennes 1, 263 avenue du g\'en\'eral Leclerc, F-35042 Rennes,
\email{nicolas.crouseilles@inria.fr}.}
\author{E. Sonnendr\"ucker}\address{Max-Planck Institute for plasma physics, Boltzmannstr. 2, D-85748 Garching,
\email{sonnen@ipp.mpg.de}.}
\author{B. Afeyan}\address{Polymath Research Inc., 827 Bonde Court, Pleasanton, CA 94566, \email{bedros@polymath-usa.com}.}

%
%
\begin{abstract} 
This work concerns
 the numerical simulation 
  of the Vlasov-Poisson equation 
using semi-Lagrangian methods on Graphical Processing Units (GPU). To accomplish this goal, modifications to traditional methods had to be implemented. First and foremost, 
 a reformulation of semi-Lagrangian methods is performed, which enables us to rewrite the governing equations as a circulant matrix operating on the vector of unknowns. 
 This product calculation can be performed efficiently using  FFT routines. Second, to overcome the limitation of single precision inherent in GPU, a $\delta f$ type 
 method is adopted which only needs refinement in specialized areas of phase space but not throughout. Thus, a GPU Vlasov-Poisson solver can indeed perform high precision  simulations (since it  uses very high order of reconstruction and a large number of grid points in phase space). We show results for more academic test cases and also for physically relevant phenomena such as the bump on tail instability and the simulation of Kinetic Electrostatic Electron Nonlinear (KEEN) waves.

\end{abstract} 

\begin{resume} 
Ce travail concerne la simulation num\'erique du mod\`ele de Vlasov-Poisson \`a l'aide de m\'ethodes 
semi-Lagrangiennes, sur des architectures GPU. Pour cela, quelques modifications de la m\'ethode 
traditionnelle ont d\^u \^etre effectu\'ees. Tout d'abord, une reformulation des m\'ethodes semi-Lagrangiennes  
est propos\'ee, qui permet de la r\'e\'ecrire sous la forme d'un produit d'une matrice circulante avec 
le vecteur des inconnues. Ce calcul peut \^etre fait efficacement gr\^ace aux routines de FFT. Puis, pour contourner 
le probl\`eme de la simple pr\'ecision, une m\'ethode de type $\delta f$ est utilis\'ee. 
Ainsi, un code Vlasov-Poisson GPU permet de simuler et de d\'ecrire avec un haut degr\'e de pr\'ecision 
(gr\^ace \`a l'utilisation de reconstructions d'ordre \'elev\'e et d'un grand nombre de points 
de l'espace des phases) des cas tests acad\'emiques mais aussi des ph\'enom\`enes physiques pertinents, comme la simulation des ondes KEEN.  
\end{resume}
\maketitle

\section*{Introduction}

 At the one body distribution function level, the kinetic theory of charged particles interacting with electrostatic fields and ignoring collisions, may be described
 by the  Vlasov-Poisson system of equations. 
This model takes into account the phase space evolution of a distribution function 
$f(t, x, v)$ where $t\geq 0$ denotes time, $x$  denotes space and $v$ is the velocity. 
Considering one-dimensional systems leads to the $1D\times 1D$ Vlasov-Poisson 
model where the solution $f(t, x, v)$ depends on   time $t\geq 0$, space $x\in [0, L]$ and velocity $v\in \mathbb{R}$. 
The distribution function $f$ satisfies 
\begin{equation}
\label{vp}
\partial_t f + v\partial_x f+E\partial_v f = 0, \;\;\;  
\end{equation}
where $E(t, x)$ is an electric field. Poisson's law dictates that the charge particle distribution must be summed over velocity to render the self-consistent electric field 
 as a solution to the  Poisson equation: 
\begin{equation}
\label{poisson}
\partial_x E = \int_{\mathbb{R}} f dv -1. 
\end{equation}
To ensure the uniqueness of the solution, we impose to the electric field a 
zero mean condition $\int_0^L E(t, x) dx = 0$. 
The Vlasov-Poisson system (\ref{vp})-(\ref{poisson})  requires an 
initial condition $f(t=0, x, v) = f_0(x, v)$.  We will restrict our attention to 
periodic boundary conditions in space and  vanishing $f$ at large velocity.   

Due to the nonlinearity of the  self-consistent evolution of two interacting fields,
in general it is difficult to find analytical solution to (\ref{vp})-(\ref{poisson}). 
This necessitates  the implementation of numerical methods to solve it. 
Historically, progress was made using particles methods (see \cite{birdsall}) 
which consist in advancing in time macro-particles through 
the equations of motion whereas the electric field is computed on a spatial mesh. 
Despite the inherent statistical numerical noise and their low convergence, 
 the computational cost of particle methods is very low
 even in higher dimensions which explains their enduring popularity. 
 On the other hand, Eulerian methods, which have been developed more recently,  
 rely on the direct gridding of phase space $(x, v)$.  Eulerian methods include  finite differences, finite volumes 
 or finite elements. Obviously, these methods are very demanding in terms 
 of memory, but can converge very fast using high order discrete operators. Among these, 
 semi-Lagrangian methods try to retain the best features of the two approaches: the phase space 
 distribution function is updated by solving backward the equations of motion 
 ({\it i.e.} the characteristics), and by using an interpolation step to remap the solution onto the phase space grid.  
 These methods are often implemented in a split-operator framework. Typically, to solve (\ref{vp})-(\ref{poisson}), 
 the strategy is to decompose the multi-dimensional problem into a sequence of $1D$ problems. 
We refer to \cite{boris, cheng, fijalkow, filbet, arber, crous, qiushu,guclu} for previous works on the subject.

The main goal of this work is to use recent GPU devices for semi-Lagrangian simulations 
of the Vlasov-Poisson system (\ref{vp})-(\ref{poisson}). 
Indeed, looking for new algorithms that are highly scalable in the field of plasmas 
 simulations (like tokamak plasmas or particle beams), it is important to mimic plasma devices more reliably.
Particle methods have already been tested on such architectures, and  good scalability 
has been obtained as in \cite{picgpu, pic-gpu}. We mention a recent precursor work 
on the parallelization in GPU in the context of a gyrokinetic eulerian code GENE 
\cite{dannert}. Semi-Lagrangian algorithms  dedicated to the simplified setting of the one-dimensionnal Vlasov-Poisson system have also 
 recently been implemented in the CUDA framework (see \cite{latu-gpu, filho}). In the latter two works, in which the interpolation step 
is based on cubic splines, one can see that the efficiency can reach a factor of  $\times 80$ in certain cases. Here, 
we use higher complexity algorithms, which are based on the Fast Fourier Transform (FFT). 
We will see that our GPU simulations will directly benefit from the huge acceleration obtained for the FFT on GPU.
 They are thus also very fast enabling us to test and compare different interpolation operators (very high order Lagrangian or spline reconstructions) 
 using a large number of grid points per direction in phase space.

To achieve this task, flexibility is required to switch easily from one representation of an operator to another. 
Here, semi-Lagrangian methods are reformulated in a framework which 
enables the use of existing optimized Fast Fourier Transform routines. This formulation gives rise to
a matrix 
which possesses the circulant property, which is a consequence of the periodic boundary conditions. 
Let us emphasize that  such boundary conditions are used not only in $x$ but also in $v$; this is made possible by taking the velocity domain 
$[-v_{\rm max},v_{\rm max}]$, with $v_{\rm max}$ big enough. Note also that the proof of convergence of such numerical schemes can be obtained
following \cite{BeMe,ChaDeMe2012}. 
 Due to the fact that such matrices are diagonalizable in a Fourier basis, the matrix 
vector product can be performed efficiently using FFT. 
In this work, 
 Lagrange polynomials of various odd degrees $(2d+1)$ and B-spline of various degree $k$ have been tested and compared.  
 Another advantage of the matrix-vector product formulation is that the numerical cost is almost  
 insensitive to the order of the method. Finally, since single precision computations are  preferable
to get maximum performance out of a GPU, other improvements have to be  made to the standard 
semi-Lagrangian method. To achieve  the accuracy needed to observe relevant physical 
phenomena, two modifications are proposed: the first is to use a $\delta f$ type method following \cite{latu-gpu}.
The second is to impose a zero spatial mean condition on the electric field.  Since the response of the plasma is periodic, this is always satisfied.

The rest of the paper is organized as follows. First, the reformulation 
of the semi-Lagrangian method using FFT is presented for the numerical treatment of 
the  doubly periodic Vlasov-Poisson model. 
Then, details of the GPU implementation are given, highlighting 
the particular modifications that were necessary in order to overcome the single precision limitation of GPUs. 
We then move on to show numerical results. These involve several comparisons between the different 
 methods and orders of numerical approximation and their performances on GPU and CPU on three canonical test problems.

\section{FFT implementation}
In this section, we give an explicit formulation of semi-Lagrangian schemes 
 for the solution of the Vlasov-Poisson system of equations in the doubly periodic case using circulant matrices. 
First, the classical directional Strang splitting (see \cite{cheng, shoucri}) is recalled. 
Then, the problem is reduced to a sequence of one-dimensional constant advections; 
Irrespective of the  method or order of the interpolation in a specific class, a circulant-matrix formulation is proposed, 
for which the use of Fast Fourier Transform is very well suited. 

\subsection{Strang-splitting}
\label{strang}
\noindent
For the  Vlasov-Poisson set of equations
 (\ref{vp})-(\ref{poisson}), it is natural to split 
the transport in the $x$-direction from
the transport in the $v$-direction. 
Moreover, this also corresponds to a splitting of the kinetic and electrostatic potential
part of the Hamiltonian $|v|^2/2 + \phi(t, x)$ where the electrostatic
potential $\phi$ is related to the electric field through $E(t, x) = -\partial_x \phi(t, x)$. 

For plasmas simulations, even when high order splittings is possible (see \cite{cfm} and references therein),  
the second order Strang splitting is a good compromise between 
accuracy and simplicity,  which explains its popularity. 
It is composed of three steps plus an update of the electric field 
before the advection in the $v$-direction
\\
\begin{enumerate}
\item Transport in $v$ over $\Delta t/2$: compute $f^{\star}(x, v)=g(\Delta t/2,x,v)$ by solving 
$$
\partial_t g(t,x,v) + E^n(x)\partial_v g(t,x,v) = 0, 
$$
with the initial condition $g(0,x,v)=f^n(x, v)$. 
\item Transport in $x$ over $\Delta t$: compute $f^{\star\star}(x, v)=g(\Delta t,x,v)$ by solving
$$
\partial_t g(t,x,v) + v\partial_x g(t,x,v) = 0, 
$$
with the initial condition $g(0,x,v)=f^{\star}(x, v)$.
\newline
Update of electric field $E^{n+1}(x)$ by solving $\partial_x E^{n+1}(x) = \int f^{\star \star}(x, v) dv -1$. 
\item Transport in $v$ over $\Delta t/2$: compute $f^{n+1}(x, v)=g(\Delta t/2,x,v)$ by solving 
$$
\partial_t g(t,x,v) + E^{n+1}(x)\partial_v g(t,x,v) = 0, 
$$
with the initial condition  $g(0,x,v)=f^{\star \star}(x, v)$. 
\end{enumerate}
One of the main advantages
of this splitting is that the algorithm reduces to solving a
series of one-dimensional constant coefficient
advections. 
Indeed, considering the transport along the $x$-direction, for each fixed $v$, 
one faces a constant advection. The same is true for the $v$-direction 
since for each fixed $x$, $E^n$ does not depend on the advected variable $v$. We choose  to start with the advection in $v$, which permits to get the electric field
 at integer multiples of time steps. The third step of the $n^{th}$ iteration could be merged with step (1) of the $(n+1)^{th}$ iteration, but we do not resort to this short cut here.

\subsection{Constant advection}
\noindent

In this part, a reformulation of  semi-Lagrangian methods is proposed, in the 
case of constant advection equations with periodic boundary conditions. 
Let us consider $u=u(t, x)$ 
to be the solution of the following equation for a given $c\in \mathbb{R}$:
$$
\partial_t u+c\partial_x u=0, \qquad u(t=0,x)=u_0(x),
$$
where periodic boundary conditions are assumed in $x\in [0, L]$. 
The continuous solution satisfies for all $t, s \geq 0$ and all $x\in[0, L]$: $u(t, x)=u(s, x-c(t-s))$. 
Let us mention that $x-c(t-s)$ has to be understood {\it modulo} $L$ since periodic boundary 
conditions are being considered.  \\ 

Let us consider a uniform mesh within
the interval $[0, L]$: $x_i=i \Delta x$ for $i=0, \dots, N$ 
and $\Delta x=L/N$. We also introduce the time step $\Delta t = t^{n+1} - t^n$ for $n\in \mathbb{N}$. 
Note that we have $u^n_0 = u^n_N$. By setting
\begin{equation}
\label{vectu}
u^{n}=\left(
  \begin{array}{c}
     u_{0}^n\\
     \vdots \\
     \vdots \\
     u_{N-1}^n\\
  \end{array} \right), 
  \quad u_i^n\approx u(t_n,x_i), 
\end{equation}
the semi-Lagrangian scheme reads
$u^{n+1}_i = \pi u^n(x_i - c\Delta t)$ where 
$\pi$ is a piecewise polynomial function which interpolates $u^n_i$ for $i=0, \dots, N-1$: 
$\pi(x_i) = u^n_i$.  
This can be reformulated into $u^{n+1} = Au^{n}$ where $A$ is the matrix defining the interpolation.
Periodic boundaries imply that the matrix $A$ is circulant:
\begin{equation}
\label{matA}
A=\mathcal{C}(a_{0},a_{1},...,a_{N-1}) :=\left(
  \begin{array}{ccccc}
     a_{0} & a_{1} & \ldots & \ldots & a_{N-1} \\
     a_{N-1} & a_{0} & a_{1} & \ldots & a_{N-2}\\
    \ddots & \ddots & \ddots & \ddots & \ddots\\
  \ddots & \ddots & \ddots & \ddots & \ddots\\
 a_{1} & \ldots & \ldots & a_{N-1} & a_{0}\\
  \end{array} \right)
\end{equation}
Obviously, this matrix depends on the choice of the polynomial reconstruction $\pi$. 
In the following, some explicit examples are shown. 

\textbf{Examples of various methods and orders of interpolation}\\
We have to evaluate $\pi u^n(x_i-c\Delta t)$. Let 
$\beta := -c \Delta t/\Delta x$ be the normalized displacement 
which can be written in a unique way as $\beta=b+b^\star$ with 
$(b,b^\star)\in \mathbb{Z} \times [0,1[$. This means that the 
feet of the characteristics $(x_i-c\Delta t)$ belong to the interval 
$[x_{i^\star}, x_{i^\star+1}[$ with $i^\star + b^\star = i+\beta$, or  
$i^\star = i+b$. 
\begin{enumerate}
\item \textit{Lagrange $1$}. The nonvanishing terms of the matrix $A$ are:
$$
a_{b}=1-b^{\star}, \qquad a_{\overline{b+1}}=b^{\star}.
$$
\item \textit{Lagrange $2d+1$} (with $2d+1\leq N-1$). The nonvanishing terms of matrix are :
$$
\forall j\in \{-d,\ldots,d+1\}, \quad a_{b+j}=\prod_{k=-d, \ k\not = j}^{d+1} \frac{b^{\star}-k}{j-k}.
$$

\item \textit{B-Spline of degree $k$}.\\
We define $B^k_i(x)$ the B-spline of degree $k$ on the mesh $(x_i)_i$ 
by the following recurrence: 
$$
B^0_i(x) = \mathds{1}_{[x_i,x_{i+1}[}(x),\qquad 
B^k_i(x)= \frac{x-x_i}{k\Delta x}B^{k-1}_i(x) +  \left(1-\frac{x-x_{i+1}}{k\Delta x}\right)B^{k-1}_{i+1}(x) .$$
Then, in this case, the nonvanishing terms of the matrix $A$ are:
$$
A=M \times \mathcal{C}(\underbrace{0,\ldots,0}_{N-k},\underbrace{B_{0}^k(x_1),B_{0}^k(x_2),\ldots,B_{0}^k(x_k)}_{k})^{-1}, 
$$
where the nonvanishing terms of the circulant matrix $M$ are:
$$
\forall j\in \{0,\ldots,k\}, \qquad m_{b-j}=B_{0}^{k}(x_{j+b^{\star}}). 
$$
\end{enumerate}

Now, starting from this reformulation, the algorithm reduces to a matrix vector product at each time step. 
Since the matrices are circulant, this product can be performed using FFT. 
Indeed, circulant matrices are diagonalizable in Fourier space \cite{gray} so that 
$$
A=UDU^{\star}, 
$$
where $U$ is unitary ($U^\star$ denotes the adjoint matrix of $U$) and $D$ is diagonal. 
They are given by
\begin{eqnarray*}
U_{m,k} &=& e^{-2i\pi mk/N}, \;\; m,k=0\ldots N-1, \\ 
D_{m,m} &=& \sum_{k=0}^{N-1}a_ke^{-2i\pi mk/N}, \;\; m=0,...,N-1. 
\end{eqnarray*}
The product of $U$ by a vector $v\in \mathbb{R}^N$ can then be obtained performing 
the Fast Fourier Transform of $v$. In the same way, $U^\star v$ can be obtained  
by computing the inverse Fourier Transform of $v$. 

The product matrix vector $Au^n=U D U^\star u^n$ is then computed following the algorithm:
\begin{enumerate}
\item Compute $U^{\star}u^n$ by calculating $\tilde{u}=\mbox{FFT}^{-1}(u^n)$.
\item Compute $D$ by calculating 
$\mbox{FFT}(a)$. 
\item Compute $w = D U^{\star}u^n$ by calculating $D \tilde{u}$.
\item Compute $A u^n$ by calculating $\mbox{FFT}(w)$.
\end{enumerate}
The complexity of the algorithm is then ${\cal O}(N\log N)$, independently of the degree 
of the polynomial reconstruction.  

\section{CUDA GPU implementation}

We use kernels on GPU by using existing NVIDIA routines for FFT, transposition and scalar product.
Note that such a choice has also been made in the more difficult context \cite{dannert}. We would have liked to use OPENCL (as done in \cite{anais}) in order not 
be attached to NVIDIA cards; but we had difficulties achieving
 the friendly well-documented features of NVIDIA, especially for the FFT.

FFTs are computed
 using the
 cufft library. For transposition, different possible algorithms are provided. The condition $N=N_x=N_v$ is always required
for this step. In order to compute charge density $\rho=\int f(t,x,v)dv$, we adapt \texttt{ScalarProd} routine.

We also write a kernel on GPU for computing coefficients of the $A$ matrix. 
An analytical formula is used for each coefficient $a_i$. In the case of Lagrange interpolation of degree $2d+1$,
the complexity switches from $O(N d)$ to $O(N d^2)$ operations because of a rewritten CPU divided differences based algorithm which cannot be parallelized.

The main steps of the algorithm are :

\begin{itemize}

\item Initialisation: the initial condition computed on CPU and transferred to GPU

\item  Computation of initial charge density $\rho$ on GPU by using \texttt{ScalarProd}

\item Transfer of $\rho$ to CPU

\item Computation of the electric field $E$ on CPU

\item Time loop
\begin{center}
\begin{itemize}

\item[1.]  $\Delta t/2$ advection in $v$ with FFT on GPU

\item[2.] Transposition in order to pass into the $x$-direction on GPU

\item[3.] $\Delta t$ advection in the $x$ direction
with FFT on GPU

\item[4.] Transposition in order to pass into the $v$-direction on GPU

\item[5.] Computation of $\rho$ on GPU by using \texttt{ScalarProd}

\item[6.] Transfer of $\rho$ to CPU

\item[7.] Computation of the electric field $E$ on CPU

\item[8.]  $\Delta t/2$ advection in $v$ with FFT on GPU

\end{itemize}
\end{center}

\end{itemize}

\section{Questions about single precision}
In principle, computations on GPU can be performed  using either single or
double precision. 
However, the numerical cost becomes quite high when one deals with double precision (we will see in our case, that the cost is generally a factor of two)
and is
 not always easily available across all platforms.
Note that in \cite{dannert} and \cite{filho}, only double precision was 
used. Discussions about single precision have already been presented in \cite{latu-gpu}.
Hereafter, we propose two slight modifications of the semi-Lagrangian method which 
enable the use of single precision computations while at the same time recovering the precision 
reached by a double precision CPU code.  

\subsection{$\delta f$ method}

The $\delta f$ method consists on a scale separation between an equilibrum and a perturbation  
so that we decompose the solution as 
$$
f(x,v) = \delta f(x,v)+f_{{\rm eq}}(v),\,\;\;  f_{{\rm eq}}(v)=\frac{1}{\sqrt{2\pi}}\exp(-v^2/2).
$$
Then, we are interested in the time evolution of $\delta f$ which satisfies 
$$
\partial_t \delta f + v\partial_x \delta f + E\partial_v  [f_{{\rm eq}} + \delta f] = 0. 
$$
The Strang splitting presented in subsection \ref{strang} is modified since we advect $\delta f$ instead of $f$. 
Since $f_{\rm eq}$ only depends on $v$, advections in $x$ are not modified. 
Now we can rewrite the $v$-advection as 
$$
\partial_t [f_{{\rm eq}} +\delta f]+ E^\star\partial_v  [f_{{\rm eq}} + \delta f] = 0, 
$$
with the initial condition $f_{{\rm eq}} +\delta f^\star$.  This means that $(f_{{\rm eq}} +\delta f)$ 
is preserved along the characteristics 
$(f_{\rm eq} + f^{\star \star})(x,v) = (f_{\rm eq} + f^{\star})(x, v-\Delta t E^\star(x))$.  
We then deduce that 
$$
\delta f^{\star \star}(x,v) = \delta f^{\star}(x,v-\Delta t E^\star(x))+ f_{{\rm eq}}(v-\Delta t E^\star(x)) - f_{{\rm eq}}(v).  
$$
which provides the update of $\delta f$ for the $v$-advection. Note that  $f_{{\rm eq}}(v-\Delta t E^\star(x))$ is an evaluation and not an interpolation.

\subsection{The zero mean condition}

The electric field is computed from (\ref{poisson}). Note that the right hand side of (\ref{poisson}) has zero mean, and the resulting electric field
has also zero mean. This is true at the continuous level; however when we deal with single precision, a systematic cumulative error could occur here.
In order to prevent this phenomenon, we can enforce the zero mean condition  on the discrete grid numerically: from 
$\rho_k^n\simeq \rho(t^n,x_k)=\int_\mathbb{R}f(t^n,x_k,v)dv,\ k=0,\dots,N-1$, we compute the mean
$$
M=\frac{1}{N}\sum_{k=0}^{N-1}\rho_k^n,
$$
and then subtract this value to $\rho_k^n$:
$$
\tilde{\rho}_k^n=\rho_k^n-M,\ k=0,\dots,N-1,
$$
so that $\tilde{\rho}_k^n\simeq \rho(t^n,x_k)-1$ is of zero mean numerically. Without this modification, we would have
$
\tilde{\rho}_k^n=\rho_k^n-1.
$
We repeat this
same procedure once the electric field is computed:
from a given computed electric field $\tilde{E}_k^n,\ k=0,\dots,N-1$, which may not be of zero mean, we compute
$
\tilde{M}=\frac{1}{N}\sum_{k=0}^{N-1}\tilde{E}_k^n,
$
and set
$$
E_k^n=\tilde{E}_k^n-\tilde{M},\ k=0,\dots,N-1.
$$

For computing the electric field, we use  the trapezoidal rule:
$$
\tilde{E}_{k+1}^n=\tilde{E}_k^n+\Delta x\frac{\tilde{\rho}_k^n+\tilde{\rho}_{k+1}^n}{2},\ k=0,\dots,N-1,
\  \textrm{with}\ \tilde{E}_0^n \ \textrm{set arbitrarily to zero},
$$
or Fourier (with FFT). Note that in the case of Fourier, the zero mean is automatically satisfied numerically, since
the mode $0$ which represents the mean is set to $0$.
$ $
\newline We will see that this zero mean condition is of great importance in the numerical results. It has to be satisfied with enough precision.
It can be viewed as being
related to the "cancellation problem" observed in PIC simulations \cite{hatzky}.
Note also, that by dealing with $\delta f,$ which is generally of small magnitude, a better resolution of the zero mean condition is reached.

\section{Numerical results}
This section is devoted to the presentation of numerical results obtained 
by the following methods: the standard semi-Lagrangian method 
(with various different
 interpolation operators), including the $\delta f$ and zero mean 
condition modifications. Comparisons between CPU and GPU 
simulations and discussions about the performance will be given 
on three test cases: Landau damping, bump on tail instability, and KEEN waves. 
As interpolation operator, we will use use by default LAG17, the Lagrange $2d+1$ interpolation with $d=8$. Similarly, LAG3
stands for $d=1$ and LAG9 for $d=4$. We will also show
simulations with standard cubic splines (for comparison purposes), which correspond to B-splines of degree $k$ with $k=4$. 
We will use several machines for GPU: MacBook, irma-gpu1 and hpc. See subsection \ref{perf} for details.

\subsection{Landau Damping}
For this first standard test case \cite{krall}, the initial condition is taken to be: 
$$
f_0(x,v)=\frac{1}{\sqrt{2\pi}}\exp\left(-\frac{v^2}{2}\right)(1+\alpha\cos(0.5 x)), \;\;\; (x,v) \in [0,4\pi] \times [-v_{\rm max},v_{\rm max}],  
$$
with $\alpha=10^{-2}$. We are interested in the time evolution 
of the electric energy ${\cal E}_e(t) = (1/2) \|E(t)\|^2_{L^2}$ 
which is known to be exponentially decreasing at
 a rate $\gamma=0.1533$ (see \cite{sonnen}). 
Due to the fact that the electric energy decreases in time, this test emphasizes the difference between 
single and double precision computations. 

Numerical results are shown on Figure \ref{landau1} (top and middle left). We use LAG17, $N=2048$, $v_{\rm max}=8$ as default values.

In the single precision case (top left), we see the benefit of using the zero mean modifications (plots 6 and 7: efft nodelta and trap zero mean nodelta):
the two results are similar (we use the trapezoidal rule
 for the electric field or Fourier and we recall that in both cases, the zero mean is satisfied)
and improved with respect to the case where the zero mean is not enforced in the trapezoidal case (plot 8: trap nodelta). 
$23$ right oscillations are reached until time $t=50$ for plots 6 and 7 (the two last oscillations are however less accurately described),  whereas we have only $16$ right oscillations until time $t=34.8$ for plot 8, before saturation. 
If we use the $\delta f$ method, we observe a further improvement (plots 1 to 5): we gain $4$ oscillations (that is we have $27$ oscillations in total) until time $t=60$,
and the electric field is below $6\cdot10^{-6}<e^{-12}$.
Note that in the case where we use the $\delta f$ method, adding the zero mean modification has no impact here; on the other hand, results with  the $\delta f$ method
are better than results with the zero mean modification on this picture. We have also added a result on an older machine (plot 9: MacBook), which leads to very poor results (only $9$ oscillations until time $t=19$ for the worst method). Also the results, which are not shown here, due to space limitations,
 were different by applying the modifications; as an example, we got $28$ right oscillations by using
the $\delta f $ method with zero mean modification. Floating point standard may not   have been satisfied there which could explain the difference in the results.

In the double precision case (top right), we can go to higher precision results.
By using $\delta f$ method or zero mean modification (the difference between both options is less visible), we get $92$ right oscillations until time $t=206$ (the last oscillation is not good resolved hat the end),
the electric field goes under $6\cdot10^{-13}<e^{-28}$, and we guess that we could add $11$ more oscillations  until time $t=231$ (we see that grid side effects polllute the result), to obtain $103$ oscillations and with electric field below $6\cdot10^{-14}<e^{-30}$, but we are limited here in the double precision case to $N=2048$.
A CPU simulation with $N=4096$ confirms the results. We also see the effect of the grid (runs with $N=1024$) and the velocity (runs with $v_{\rm max}=6$).
Note that the plot 6 (trap nodelta 1024 v6) has lost a lot of accuracy compared to the other plots: the grid size is too small ($N=1024$), the velocity domain also
($v_{\rm max}=6$) and above all there is no zero mean or $\delta f$ method. In that case, we only reach time $t=100$. 
We refer to \cite{zhou,filbet_landau} for other numerical results and discussions and to the seminal famous work \cite{villani} for theoretical results.  In \cite{zhou}, it was already mentionned that we have to take the velocity domain large enough
and to take enough grid points. Concerning GPU and single precision, the benefit of a $\delta f$ method was also already treated in \cite{latu-gpu}: $29$ right oscillations
were obtained in the single precision case with a $\delta f$ modification, $13$ right oscillations without the modification and the time $t=100$ was reached in the CPU case ($N$ was set to $1024$ and 
$v_{\rm max}$ to $6$).

On Figure \ref{landau1} middle left, we plot the error of mass, which is computed as $|\hat{\rho}_0-1|$. We clearly see the impact between the conservation of the mass and the former results.
We can also note that, the zero mean modification does not really improve the mass conservation (just a slight improvement at the end, plots 2,3,4), but has a benefic effect
on the electric field: the bad behaviour of the mass conservation is not propagated to the electric field. On the other hand, the $\delta f$ method clearly improves the mass conservation.
We also see the effect of taking a too small velocity domain, in the double precision case. 

\subsection{Bump on tail}
\label{subsection_bot}
For this second standard test case, the initial condition is considered as a spatial and periodic 
perturbation of two Maxwellians  (see \cite{shoucri}) 
$$
f_0(x,v)=\left( \frac{9}{10 \sqrt{2 \pi}} \exp\left(-\frac{v^2}{2}\right)+\frac{2}{10 \sqrt{2 \pi}} \exp(-2(v-4.5)^2) \right)(1+0.03 \cos(0.3x)), \;\;\; (x,v) \in [0,20\pi] \times [-9,9].  
$$
The Vlasov-Poisson model preserves some physical quantities with time, called Casimir functions, which will be used 
to compare the different implementations. Particulary, we look at the time history of the total energy 
${\cal E}$ of the system, which is the sum of the kinetic energy ${\cal E}_k$ 
and the electric energy  ${\cal E}_e$
$$
{\cal E}(t) ={\cal E}_k(t)+{\cal E}_e(t)= \int_0^{4\pi}\int_{\mathbb{R}} f(t, x, v) \frac{v^2}{2} dvdx + \frac{1}{2}\int_0^{4\pi} E^2(t, x) dx. 
$$
As in the previous case, the time evolution of the electric energy is chosen as a diagnostics. 

\begin{figure}[h!]
\begin{tabular}{c}
\includegraphics[width=0.5\linewidth,angle=0,height=6cm]{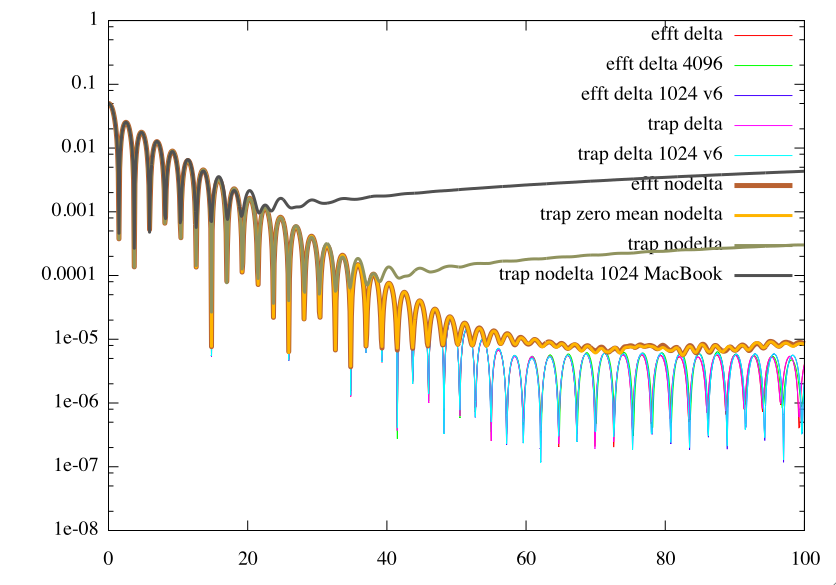}
\includegraphics[width=0.5\linewidth,angle=0,height=6cm]{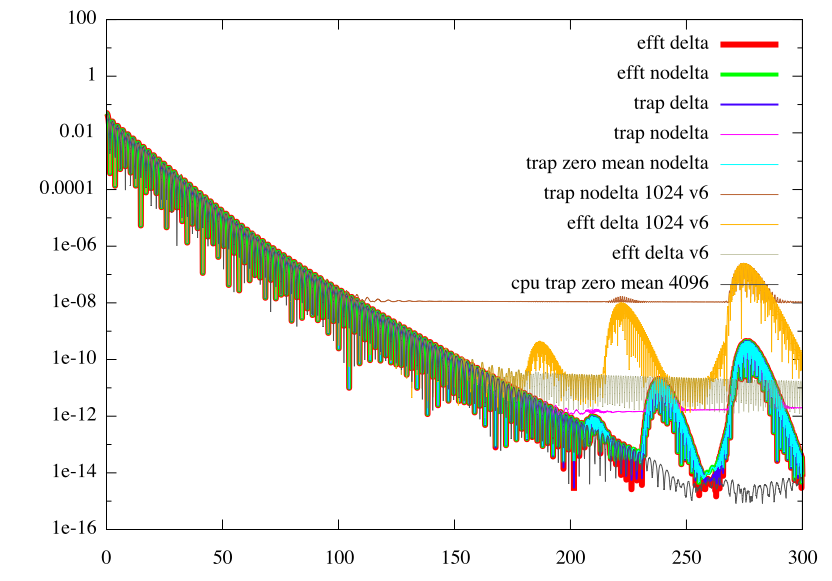} \\
\includegraphics[width=0.5\linewidth,angle=0,height=6cm]{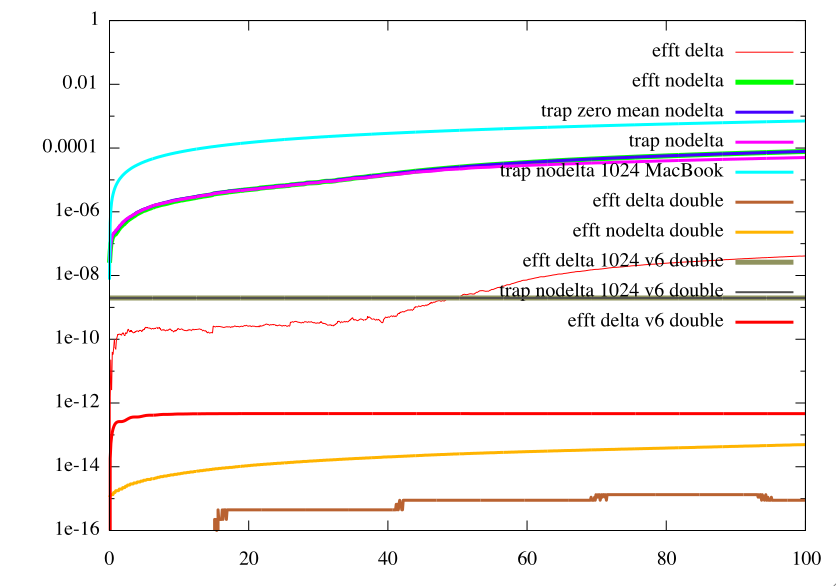}
\includegraphics[width=0.5\linewidth,angle=0,height=6cm]{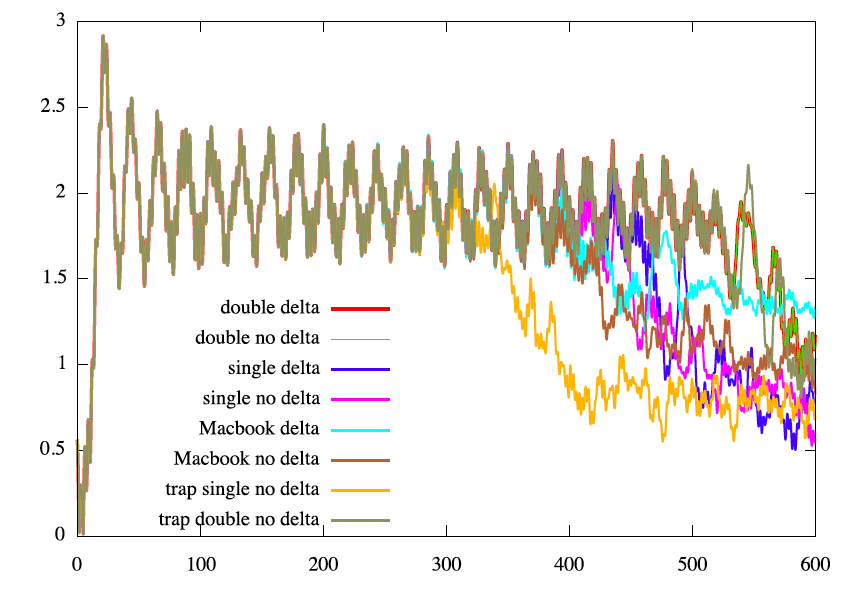} \\
\includegraphics[width=0.5\linewidth,angle=0,height=6cm]{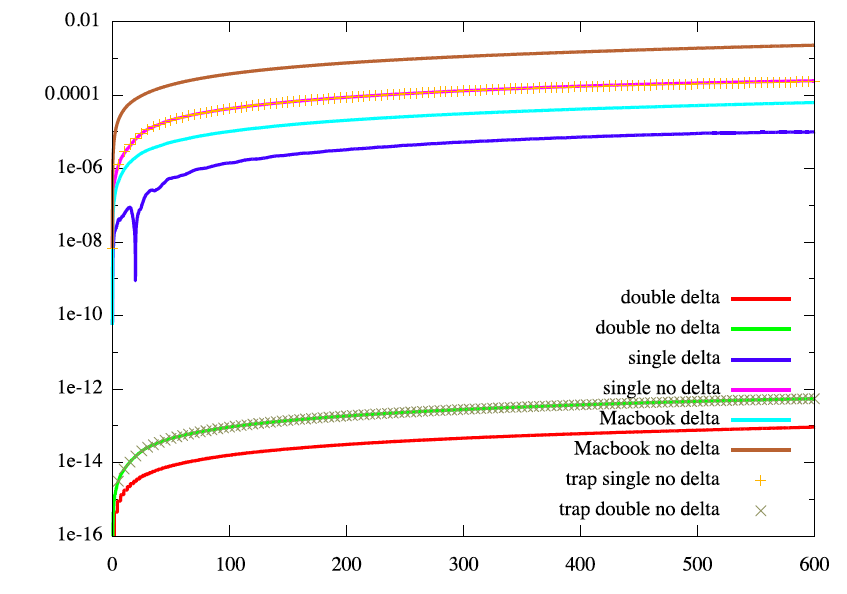}
\includegraphics[width=0.5\linewidth,angle=0,height=6cm]{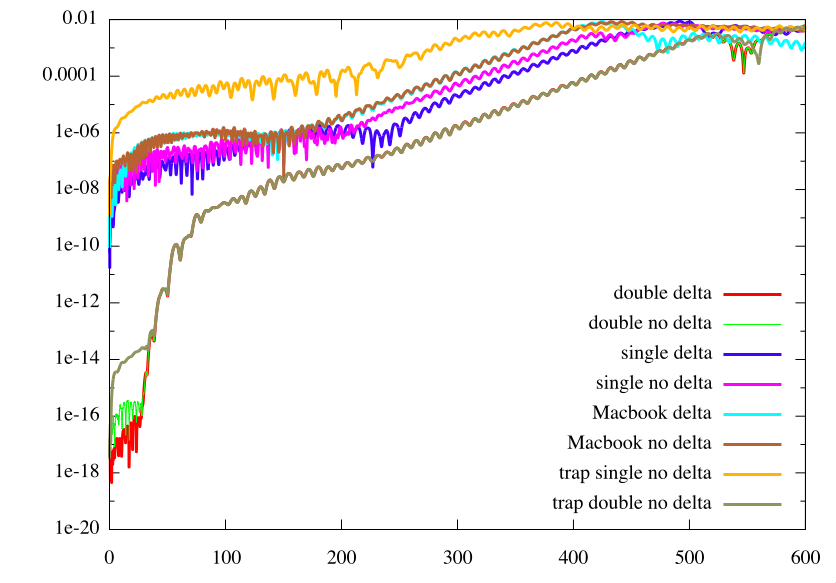} \\
\end{tabular}
\caption{
Linear Landau damping. $N=2048,\ \Delta t =0.1$, $v_{\max}=8$, LAG17, irma-gpu1 on GPU as default.
Evolution in time of electric energy in single/double precision (top left/right).
Error of mass $|\hat{\rho}_0-1|$ with single precision as default (middle left).
Bump on tail test case. $N=1024, \Delta t =0.05$,  LAG9,  irma-gpu1 on GPU as default.
Evolution in time of the electric energy/ error of mass/ first Fourier mode of $\rho$, $|\hat{\rho}_1|$ (middle right/bottom left/bottom right).
[ for details, see the legends. efft:electric field compute with FFT; delta= $\delta f$ method; no delta= without the $\delta f$ method;
single=single precision; double:double precision; trap:electric field computed with trapezoidal method;
zero mean:zero mean modification for the electric field; cpu: cpu code used; 1024: $N=1024$; 4096: $N=4096$; v6: $v_{\rm max}=6$; Macbook: MacBook GPU is used].
}
\label{landau1}
\end{figure}

\begin{figure}[h!]
\begin{tabular}{c}
\includegraphics[width=0.5\linewidth,angle=0,height=6cm]{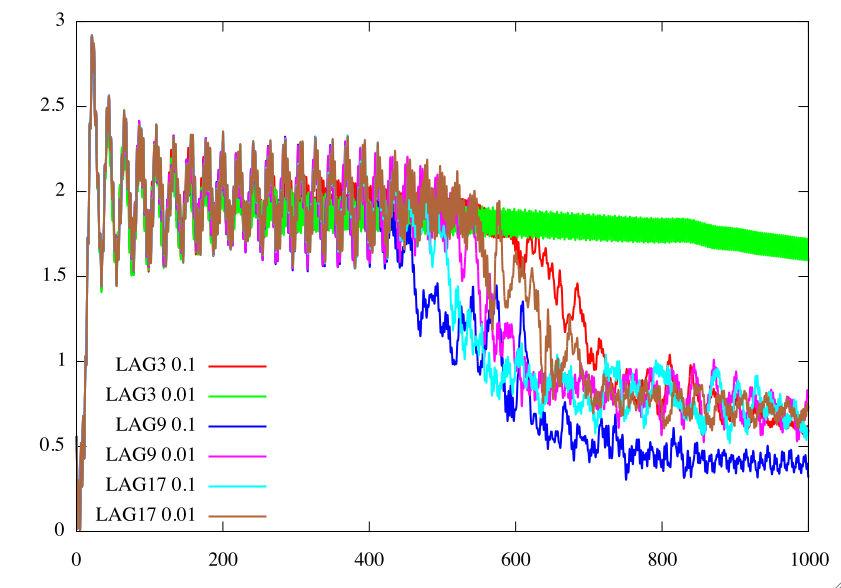}
\includegraphics[width=0.5\linewidth,angle=0,height=6cm]{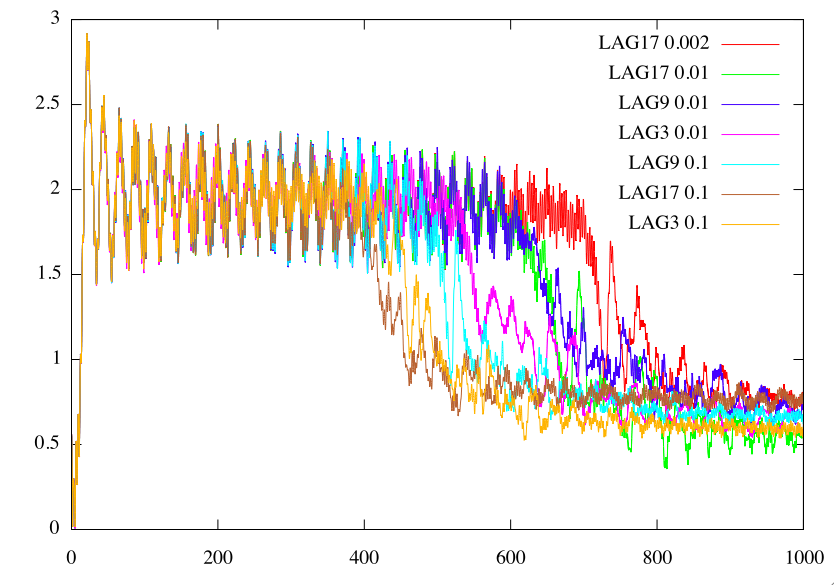} \\
\includegraphics[width=0.5\linewidth,angle=0,height=6cm]{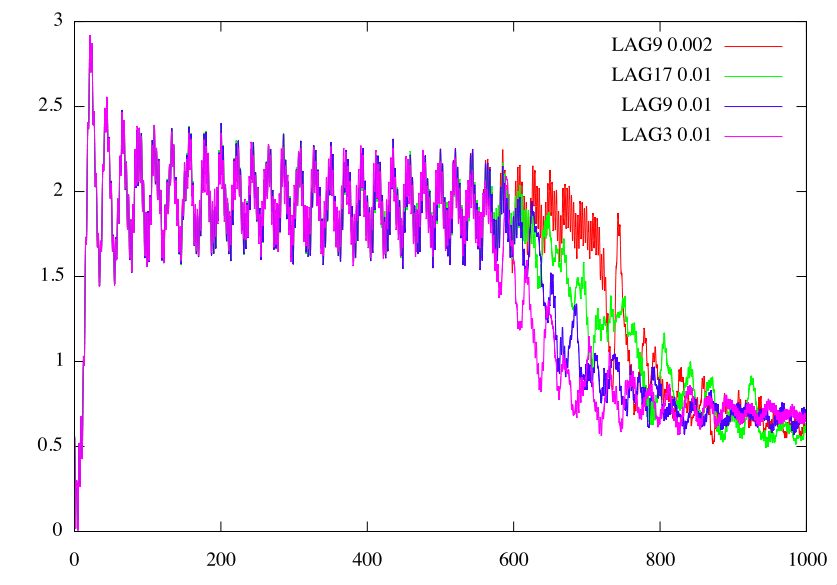}
\includegraphics[width=0.5\linewidth,angle=0,height=6cm]{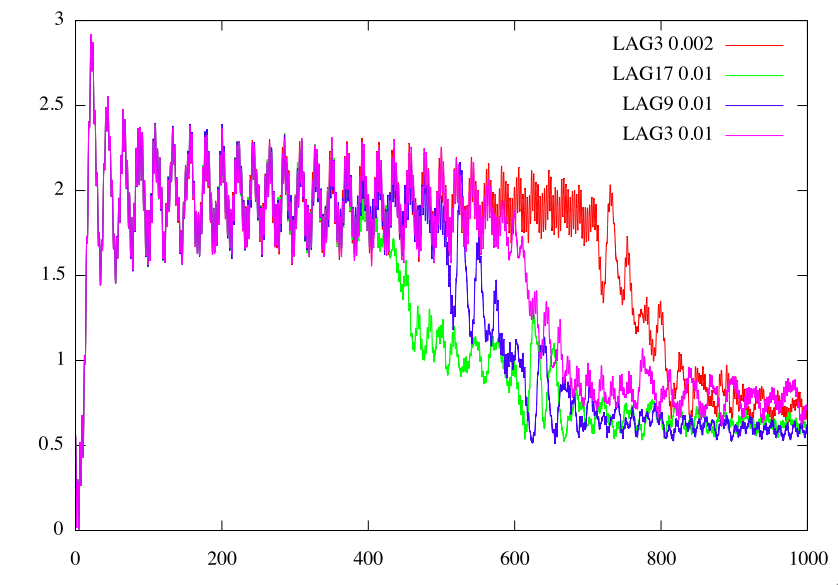} \\
\includegraphics[width=0.5\linewidth,angle=0,height=6cm]{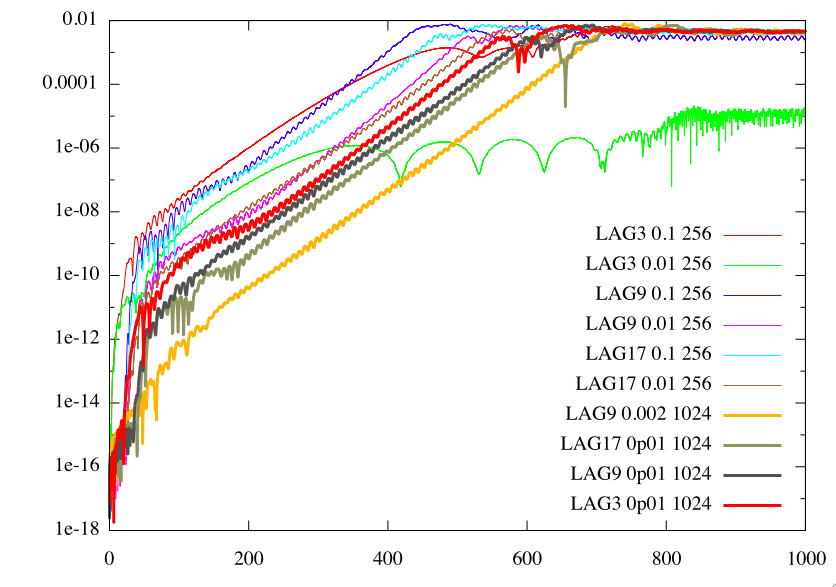}
\includegraphics[width=0.5\linewidth,angle=0,height=6cm]{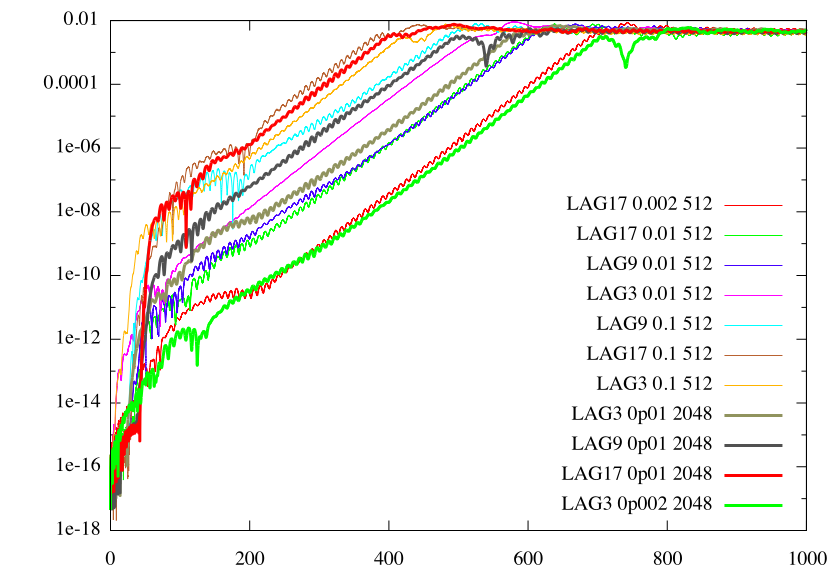} \\
\end{tabular}
\caption{Bump on tail test case. Double precision is used, irma-gpu1 on GPU.
Evolution in time of the electric energy for $N=256,512,1024,2048$ (top left, top right, middle left, middle right),
with LAG3, LAG9, LAG17 recosntructions and various time steps ($0.1,0.01,0.002$).
Evolution in time of the first Fourier mode , $|\hat{\rho}_1|$ for $N=256$ and $N=1024$ (bottom left), and for $N=512$ and $N=2048$ (bottom left),
with the same reconstructions and time steps.
}
\label{bot1}
\end{figure}

Results are shown on Figure \ref{landau1} (middle right and bottom) and on Figure \ref{bot1}.

We see on Figure \ref{landau1} middle right the evolution in time of the electric field. Single and double precision results are compared.
In the single precision case, the $\delta f$ method with FFT computation of the electric field (plot 3: single delta) is the winner
and the basic method without modifications and trapezoidal computation of the electric field (plot 7: trap single no delta) leads to the worst result.
Double precision computations lead to better results and differences are small: plots 1 (double delta) and 2 (double no delta) are undistinguishable
and plot 8 (trap double no delta) is only different at the end. Thus, such modifications are not so  mandatory in the double precision case.
We then see for the same runs, the evolution of the error of mass (bottom left) and of the first mode of $\rho$ in absolute value (bottom right). 
We notice that the error of mass linearly accumulates in time. Here no error coming from the velocity domain is seen, because $v_{\rm max}$ is large enough ($v_{\rm max}=9$ in all the runs).
The evolution of the first mode of $\rho$ is quite instructive: we see that it exponentially grows from round off errors and the different runs lead to quite different results.
The loss of mass can become critical in the single precision case (no real impact in the double precision case are detected)
and implementations without mass error accumulation would be desirable. The $\delta f$ method improves the results, but the error of mass still accumulates
 much more than in the double precision case. 
 
On Figure \ref{bot1}, we see the same diagnostics in the double precision case. We make vary the number of grid points, the degree of the interpolation and the time step. 
By taking smaller time step, we can increase the time before the merge of two vortices among three which leads to a breakdown of the electric field.
Higher degree interpolation lead to better results (in the sense that the breakdown occurs later), for not too high grid resolution. 
When $N=2048$, lower order interpolation seems to be better, since it introduces more diffusion, whereas high order schemes try to capture the small scales, which are then sharper
and more difficult to deal with in the long run. Adaptive methods and methods with low round-off error in the single precision case may be helpful to get better results.

\subsection{KEEN Waves}
 In this last and most intricate test, instead of considering a perturbation of the initial data, 
 we add an external driving electric field $E_{\rm app}$ to the Vlasov-Poisson equations:
\begin{eqnarray*}
\partial_t f + v\partial_x f+(E-E_{\rm app})\partial_v f = 0,\ \partial_x E = \int_{\mathbb{R}} f dv -1, 
\end{eqnarray*}
where $E_{\rm app}(t, x)$ is of the form
$
E_{\rm app}(t,x)=E_{\rm max}ka(t)\sin(kx-\omega t), 
$
where $$
a(t)=\frac{0.5(\tanh(\frac{t-t_L}{t_{wL}}) - \tanh(\frac{t-t_R}{t_{wR}}))- \epsilon}{1-\epsilon},\ 
 \epsilon = 0.5(\tanh(\frac{t_0-t_L}{t_{wL}}) - \tanh(\frac{t_0-t_R}{t_{wR}}))
$$ is the amplitude, $t_0=0,\ t_L=69,\ t_R=307,\ t_{wL}=t_{wR}=20$,
 $k=0.26$, $\omega=0.37$ and $E_{\rm max}=0.2$.
The initial condition is 
$$
f_0(x, v) = \frac{1}{\sqrt{2\pi}}\exp\left(-\frac{v^2}{2}\right), \;\;\; (x,v) \in [0,2\pi/k] \times [-6,6].   
$$
See \cite{bedros,keen_sonnen} for details about this physical test case. Its importance stems from the fact that KEEN waves represent new non stationary multimode oscillations of nonlinear kinetic plasmas with no fluid limit and no linear limit. They are states of plasma self-organization that do not resemble the (single mode) way in which the waves are initiated. At low amplitude, they would not be able to form.  KEEN waves can not exist off the dispersion curves of classical small amplitude waves unless a self-sustaining vortical structure is created in phase space, and enough particles trapped therein, to maintain  the self-consistent field, long after the drive field has been turned off.   For an alternate method of numerically simulating the Vlasov-Poisson system using the discontinuous Galerkin approximation, see  \cite{cgm2012} for a KEEN wave test case.

\noindent As diagnostics, we consider here different snapshots of $f-f_0$ at different times: $t=200, t=300, t=400, t=600$ and $t=1000$.

\begin{figure}[h!]
\begin{tabular}{c}
\includegraphics[width=0.5\linewidth,angle=0,height=6cm]{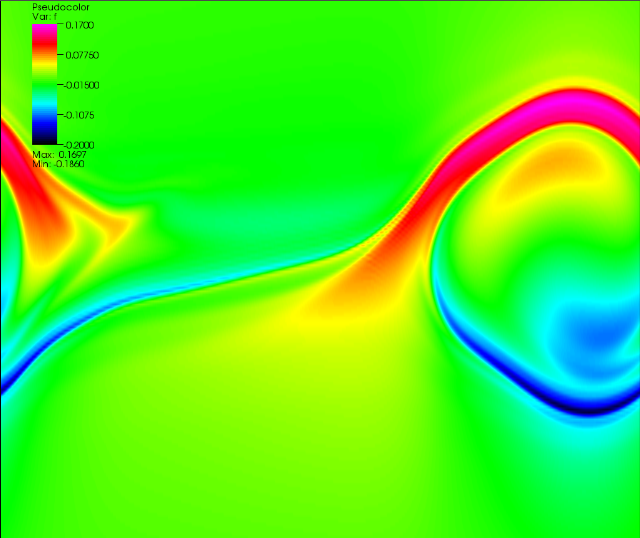}
\includegraphics[width=0.5\linewidth,angle=0,height=6cm]{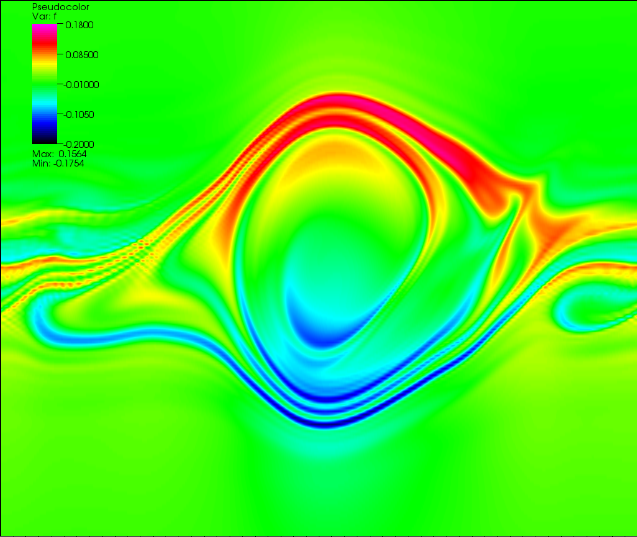} 
\\
\includegraphics[width=0.5\linewidth,angle=0,height=6cm]{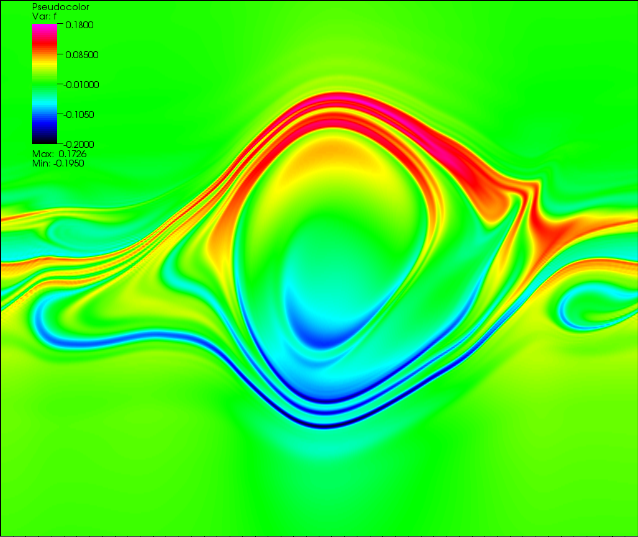} 
\includegraphics[width=0.5\linewidth,angle=0,height=6cm]{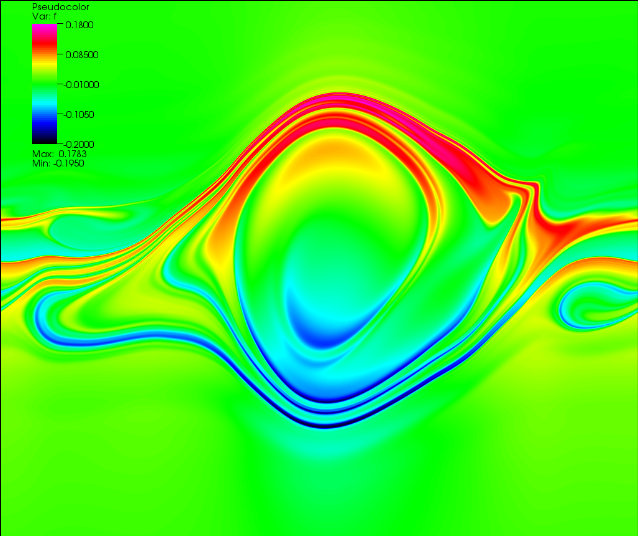} 
\\
\includegraphics[width=0.5\linewidth,angle=0,height=6cm]{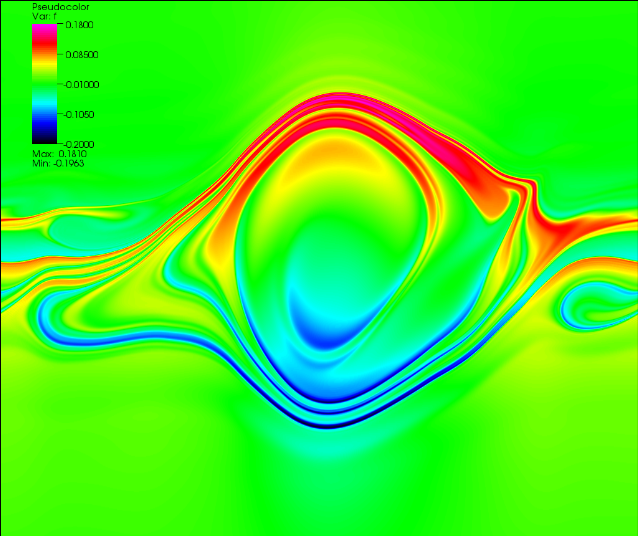} 
\includegraphics[width=0.5\linewidth,angle=0,height=6cm]{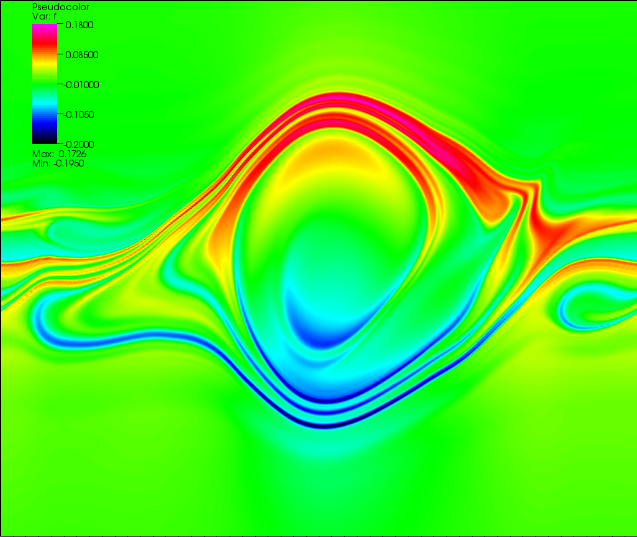}  
\end{tabular}
\caption{KEEN wave test case (LAG17): $f(t,x,v)-f_0(x,v)$. At time $t=200$, GPU single precision $N=1024$ (top left). 
At time $t=300$, GPU single precision $N=1024, 2048, 4096$ 
and $\Delta t=0.1$ (top right, middle left, middle right). $N=4096$ and $\Delta t=0.01$ (bottom left). 
CPU $N=2048,\Delta t=0.1$ (bottom right).
$(x,v)\in [0,2\pi/k]\times[0.18,4.14]$. If not changed, from one picture to another (from top left to bottom right), parameters are not restated.}
\label{keen1}
\end{figure}

We first consider the time $t=200$ (upper left in Figure \ref{keen1}). At this time, all the
snapshots are similar so we present only one (GPU single precision and a grid of $1024^2$ points). The five others graphics of this figure are taken at time t=300. We show that, at this time, there is again convergence because the graphic on the middle right (GPU single precision, $N=4096$ and $\Delta t=0.1$) is identical to the bottom left one (GPU single precision, $N=4096$ and $\Delta t=0.01$).

\begin{figure}[h!]
\begin{tabular}{c}
\includegraphics[width=0.5\linewidth,angle=0,height=6cm]{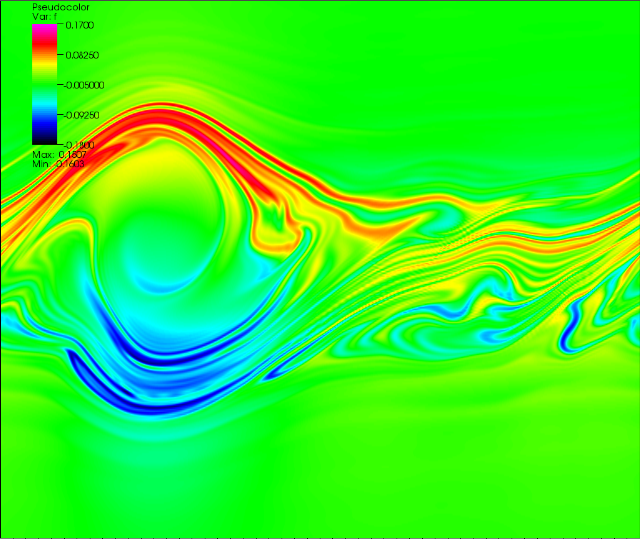} 
\includegraphics[width=0.5\linewidth,angle=0,height=6cm]{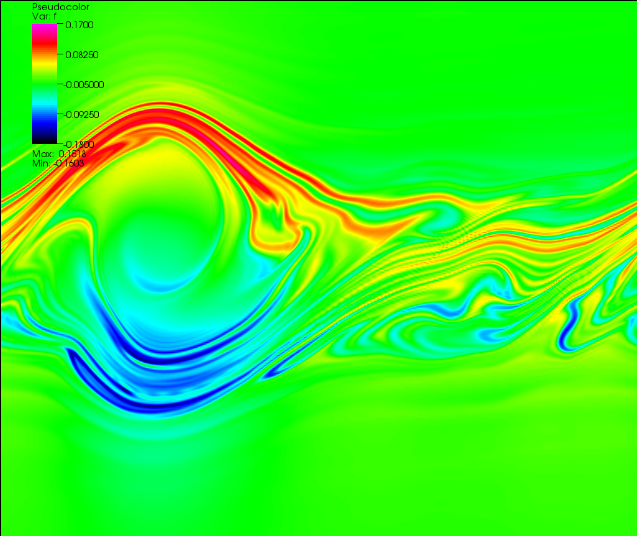}
\\
\includegraphics[width=0.5\linewidth,angle=0,height=6cm]{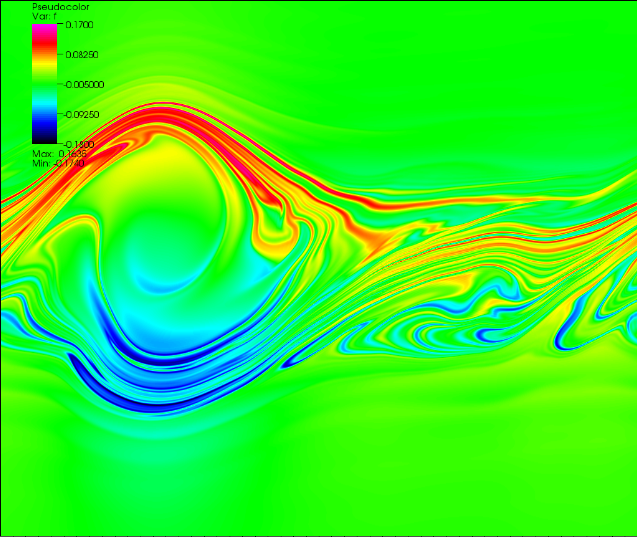}
\includegraphics[width=0.5\linewidth,angle=0,height=6cm]{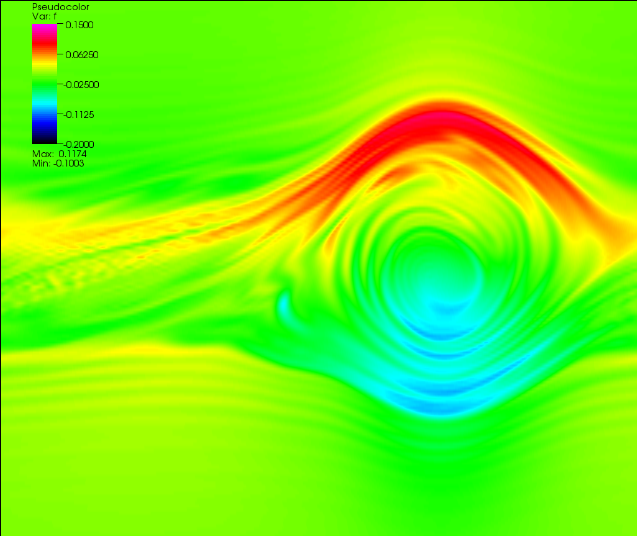}
\\
\includegraphics[width=0.5\linewidth,angle=0,height=6cm]{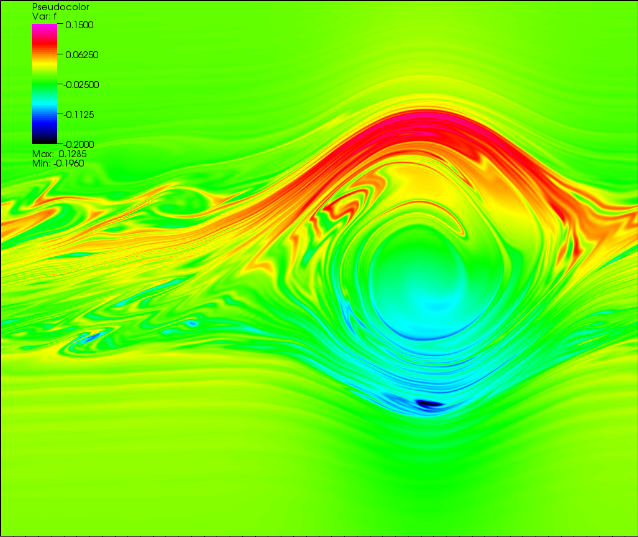} 
\includegraphics[width=0.5\linewidth,angle=0,height=6cm]{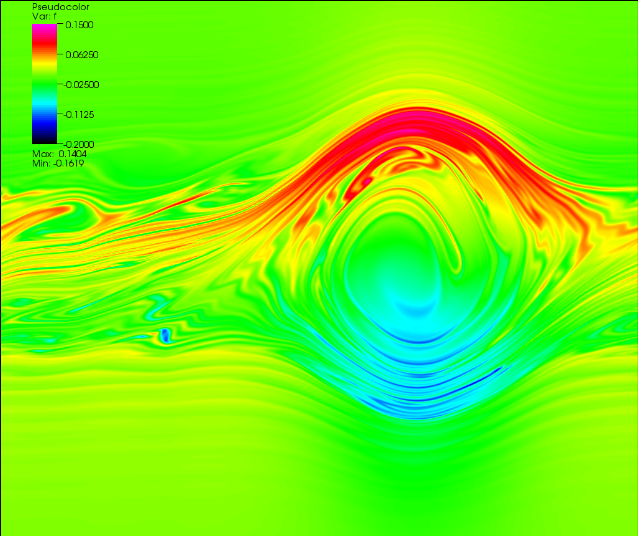}
\end{tabular}
\caption{KEEN wave test case (LAG17): $f(t,x,v)-f_0(x,v)$. At time $t=400$, GPU single precision $N=2048, \Delta t =0.1$ (top left).
CPU $\Delta t=0.05$ (top right). GPU single precision $N=4096,\Delta t=0.1$, at time $t=600$ (middle left). GPU single precision $N=1024$ (middle right).
$\Delta t=0.01, N=4096$ (bottom left). CPU $N_x=512, N_v=4096$ (bottom right). $(x,v)\in [0,2\pi/k]\times[0.18,4.14]$.
If not changed, from one picture to another (from top left to bottom right), parameters are not restated.}
\label{keen2}
\end{figure}

The Figure \ref{keen2} presents different snapshots at times $t=400$ and $t=600$. At time $t=400$, the upper left graphic (GPU single precision, $N=2048$, $\Delta t=0.1$) is similar to the upper right one (CPU, $N=2048$, $\Delta t=0.05$), that shows that the CPU and the GPU codes give the same results. With 4096 points (on middle left), we observe a little difference with the 2048 points case. Between the snapshots at time $t=400$ and those at time $t=600$, we observe the emergence of diffusion. 

\begin{figure}[h!]
\begin{tabular}{c}
\includegraphics[width=0.5\linewidth,angle=0,height=6cm]{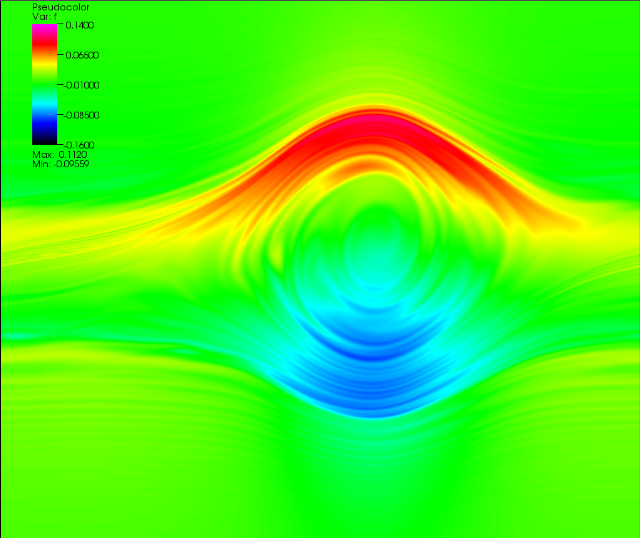}
\includegraphics[width=0.5\linewidth,angle=0,height=6cm]{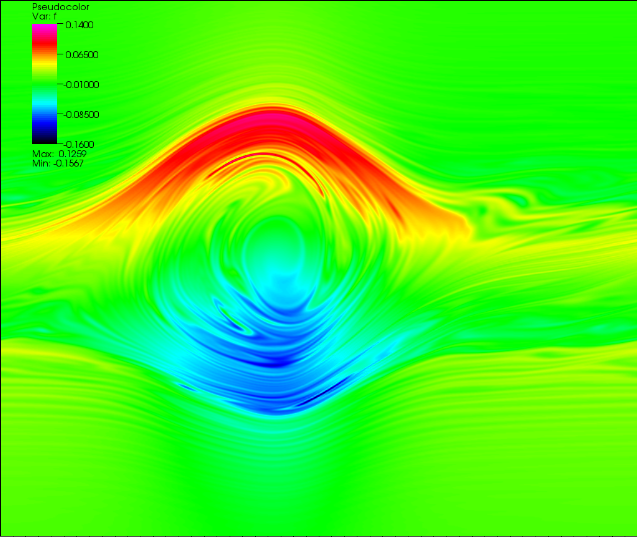}
\\
\includegraphics[width=0.5\linewidth,angle=0,height=6cm]{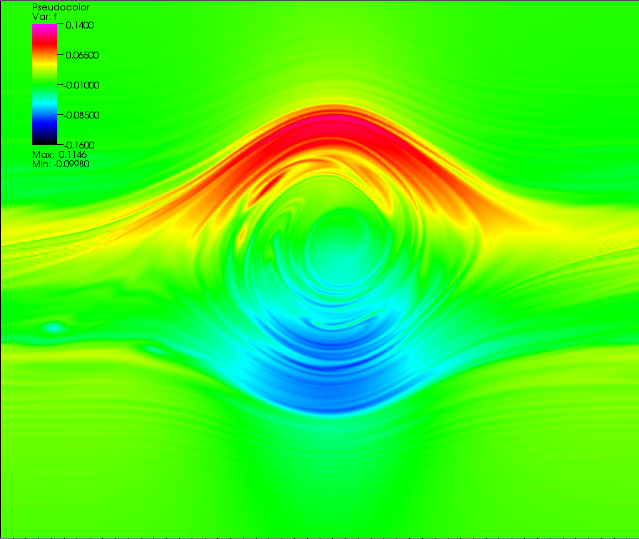} 
\includegraphics[width=0.5\linewidth,angle=0,height=6cm]{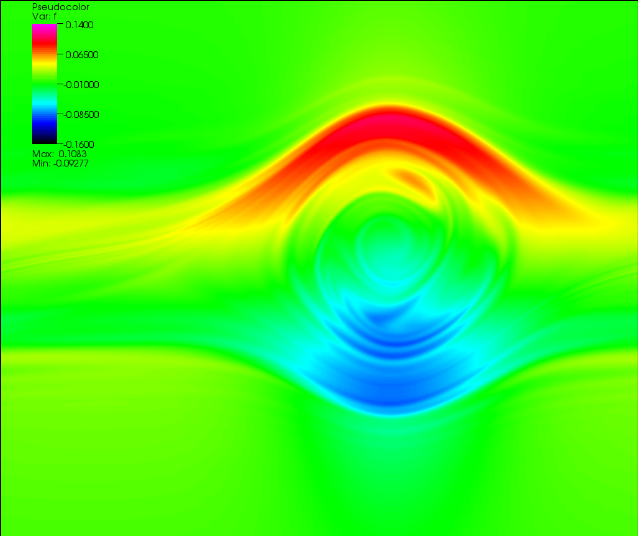} 
\\
\includegraphics[width=0.5\linewidth,angle=0,height=6cm]{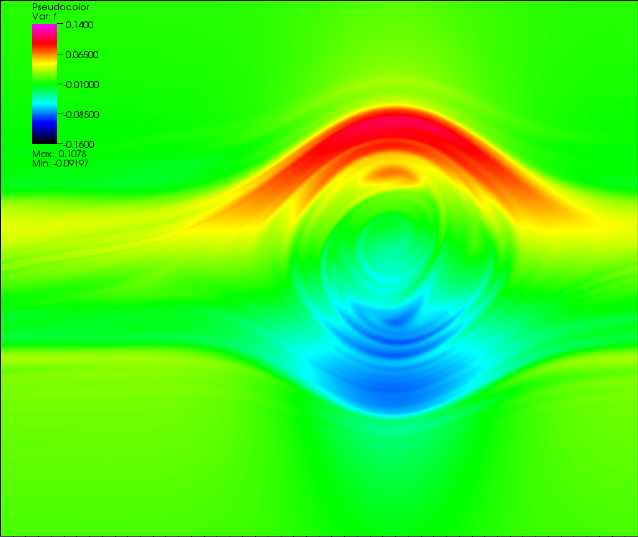} 
\includegraphics[width=0.5\linewidth,angle=0,height=6cm]{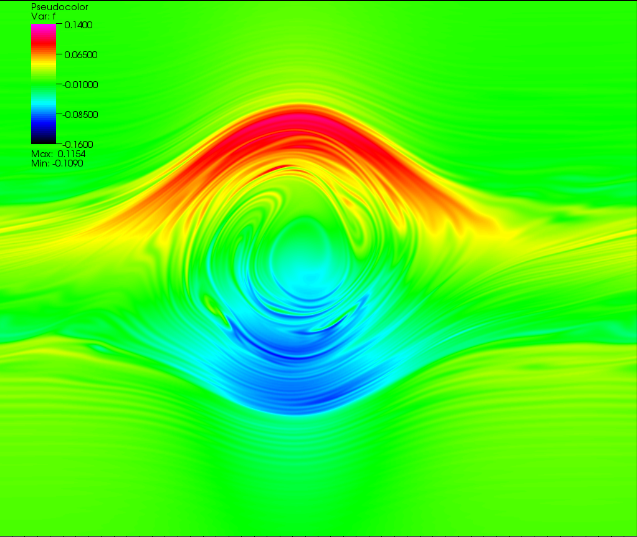} 
\end{tabular}
\caption{KEEN wave test case: $f(t,x,v)-f_0(x,v)$ at time $t=1000$. CPU cubic splines, $\Delta t=0.05, N_x=512, N_v=4096$ (top left). 
LAG17 (top right). $N=4096$ and cubic splines (middle left). LAG3 (middle right). 
GPU single precision (top left). LAG9 (bottom right). $(x,v)\in [0,2\pi/k]\times[0.18,4.14]$.
If not changed, from one picture to another (from top left to bottom right), parameters are not restated.}
\label{keen3}
\end{figure}

The time $t=1000$ is considered on Figure \ref{keen3}. We see that there is no more convergence at this time: there is a lag, but the structure remains the same. We compare also different interpolators (cubic splines, LAG 3, LAG 9, LAG 17). If the order of the interpolation is high (graphic at the top right : CPU, LAG 17, $\Delta t=0.05, N_x=512, N_v=4096$) there is appearance of finer structures. At this time, one sees little difference between CPU results (graphic at the middle right : CPU, LAG 3, $\Delta t=0.05, N=4096$) and GPU results (graphic at the bottom left : GPU, LAG 3, $\Delta t=0.05, N=4096$), but there is no lag.

\begin{figure}[h!]
\begin{tabular}{c}
\includegraphics[width=0.5\linewidth,angle=0,height=6cm]{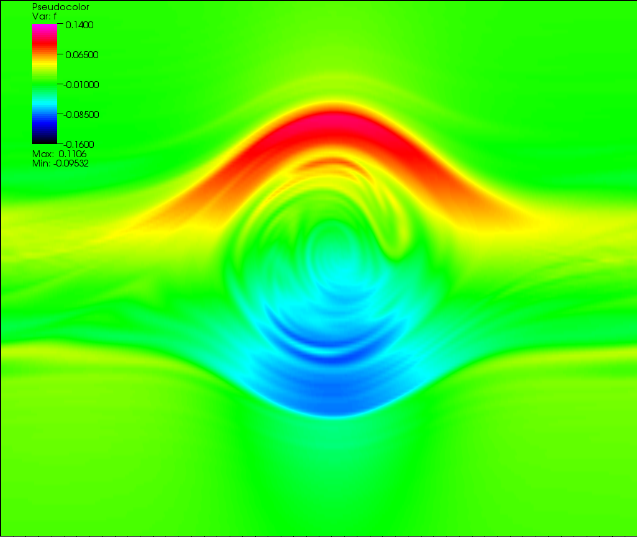} 
\includegraphics[width=0.5\linewidth,angle=0,height=6cm]{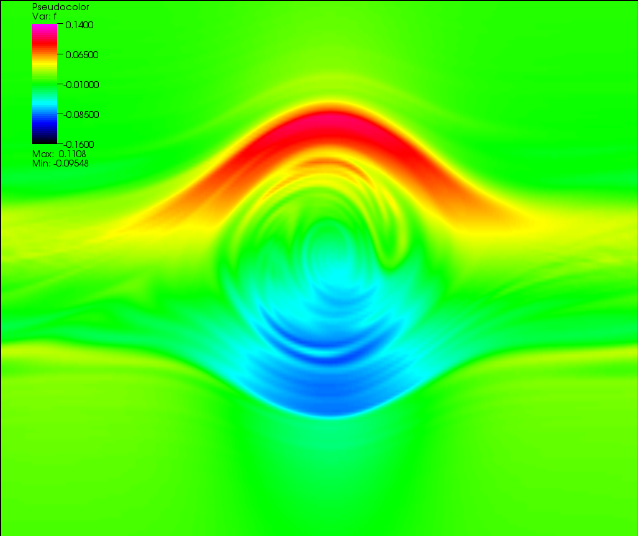}
\\
\includegraphics[width=0.5\linewidth,angle=0,height=6cm]{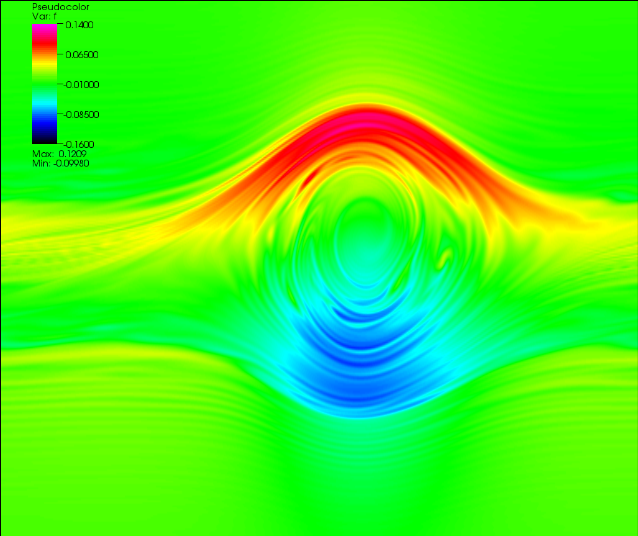} 
\includegraphics[width=0.5\linewidth,angle=0,height=6cm]{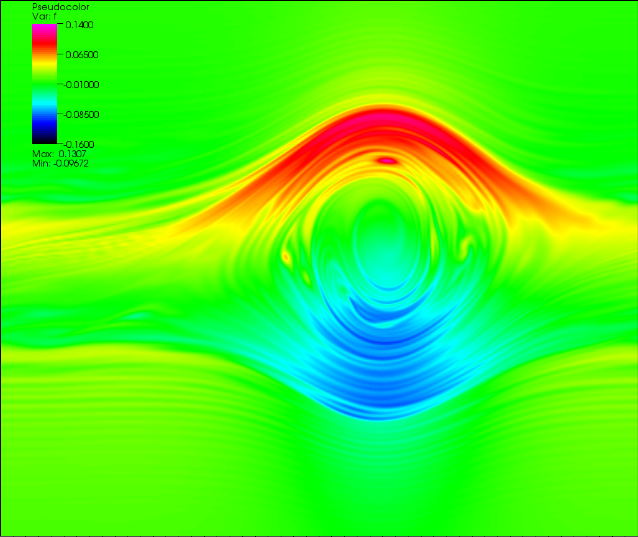} 
\\
\includegraphics[width=0.5\linewidth,angle=0,height=6cm]{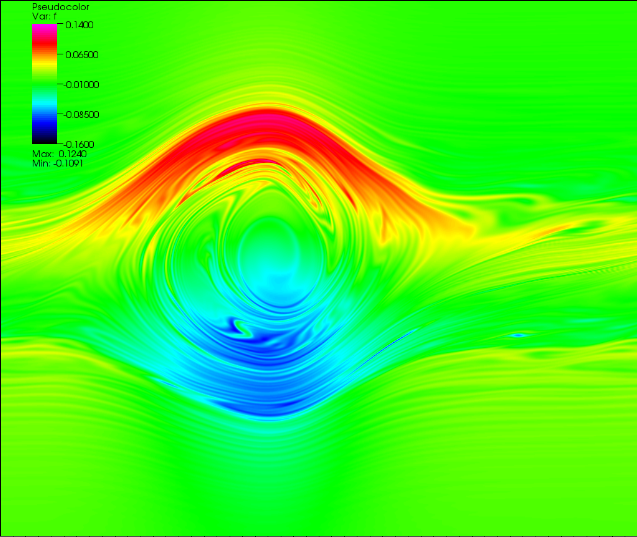} 
\includegraphics[width=0.5\linewidth,angle=0,height=6cm]{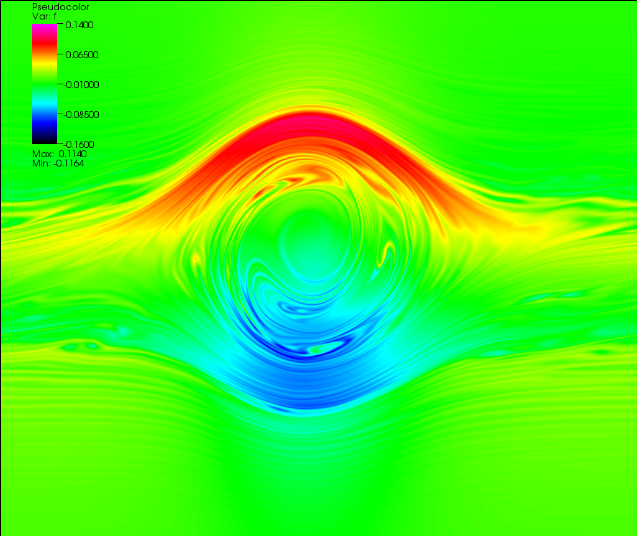} 
\end{tabular}
\caption{KEEN wave test case (LAG17): $f(t,x,v)-f_0(x,v)$ at time $t=1000$, $\Delta t=0.05, N=1024$ as default. GPU single/double precision (top left/right).
$N=2048$, GPU single/double precision (middle left/right). $N=4096$, GPU single precision (bottom left). $\Delta t=0.01$ (bottom right).
$(x,v)\in [0,2\pi/k]\times[0.18,4.14]$. If not changed, from one picture to another (from top left to bottom right), parameters are not restated.}
\label{keen4}
\end{figure}

The Figure \ref{keen4} (at time $t=1000$) shows the differences between single and double precision when the value of $N$ is changed. The two graphs above show the case $N = 1024$, the left is single precision while the right one is in double precision. We see that there are very few differences. When $N = 2048$, the results are different in single precision (graphic on middle left) and double precision (graphic on middle right). When $N = 4096$, the code does not work in double precision so we compared the results for single precision GPU with $\Delta t =0.05$ (bottom left graphic) and $\Delta t=0.01$ (bottom right graphic). There are also differences due to the non-convergence. Moreover, we see that there are more filamentations when $N$ increases.

\begin{figure}[h!]
\begin{tabular}{c}
\includegraphics[width=0.5\linewidth,angle=0,height=6cm]{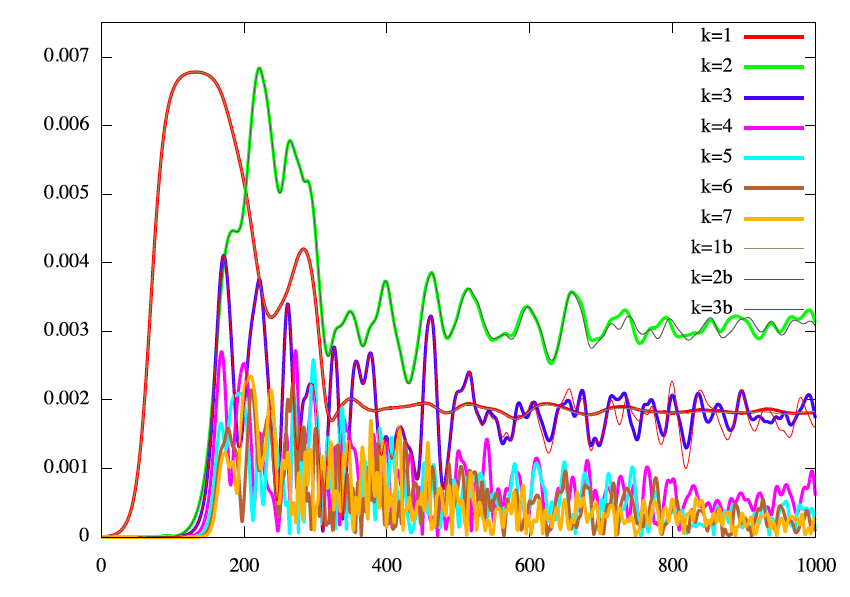} 
\includegraphics[width=0.5\linewidth,angle=0,height=6cm]{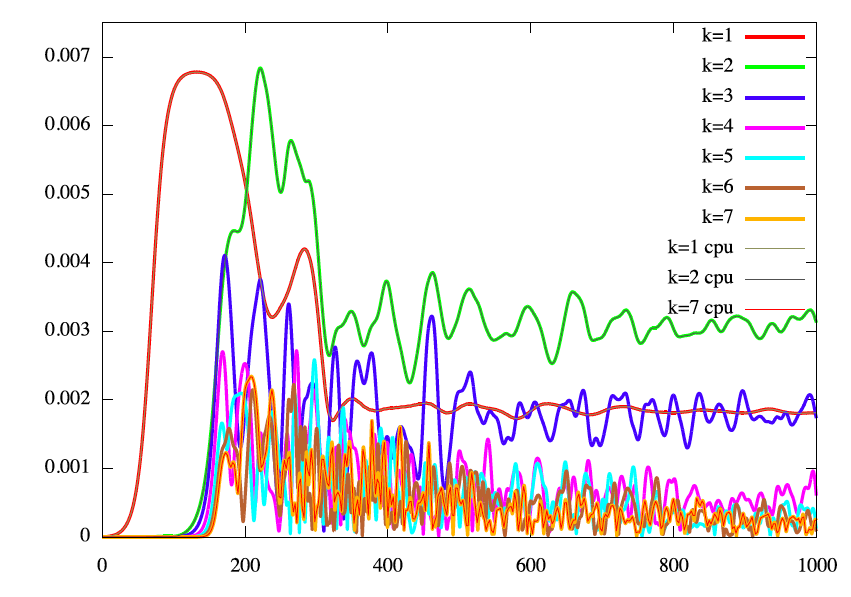}
\\
\includegraphics[width=0.5\linewidth,angle=0,height=6cm]{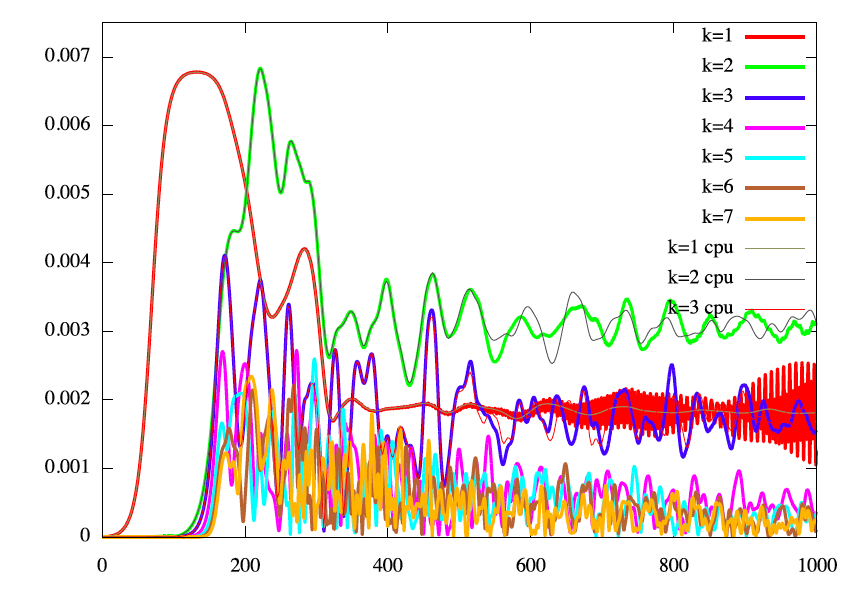} 
\includegraphics[width=0.5\linewidth,angle=0,height=6cm]{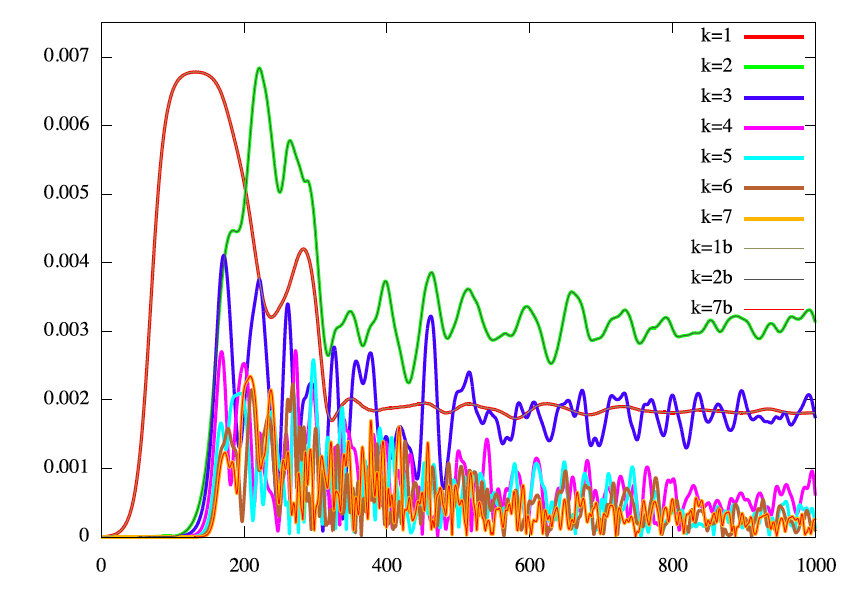} 
\\
\includegraphics[width=0.5\linewidth,angle=0,height=6cm]{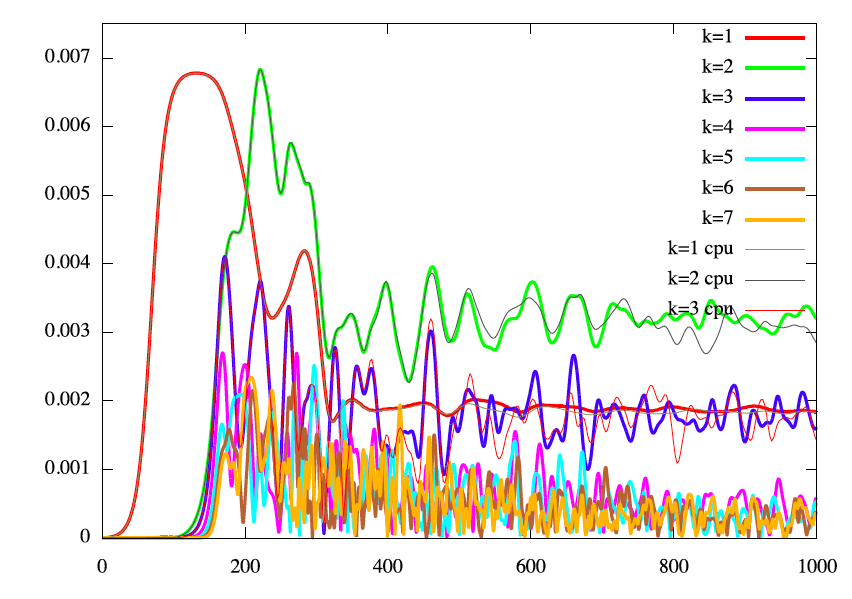} 
\includegraphics[width=0.5\linewidth,angle=0,height=6cm]{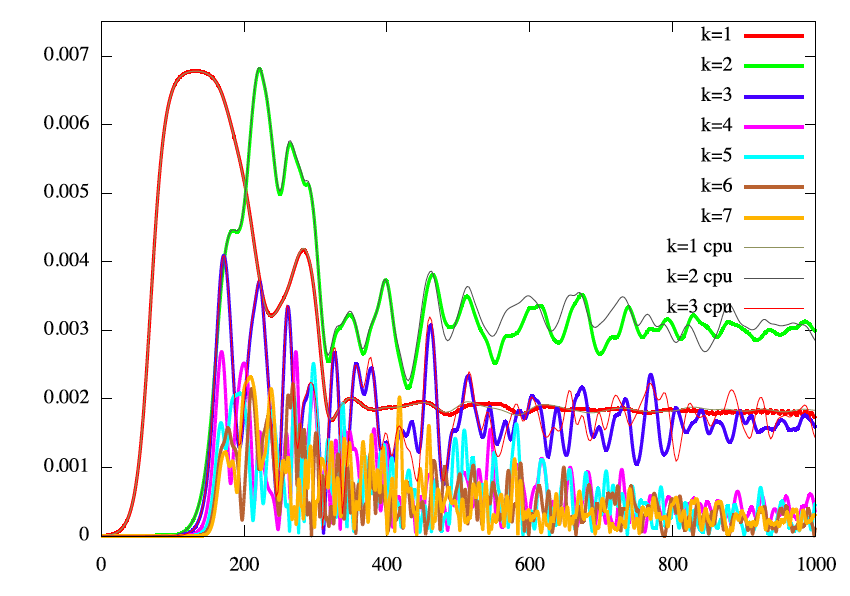} 
\end{tabular}
\caption{KEEN wave test case (LAG17): Absolute values of the first Fourier modes of $\rho$ (from mode $k=1$ to mode $k=7$) vs time.
$\delta f$ method, with $N=2048$ $\Delta t=0.05$ GPU, double and single precision (1b,2b,3b) (top left).
double GPU and double CPU (top right). Full version GPU in single precision and $\delta f$ version CPU (middle left).
Full version and $\delta f$ version, in double precision (middle right).
$N=4096$, GPU and CPU (bottom left). GPU with $\Delta t =0.01$ and CPU with $\Delta t =0.05$ (bottom right). If not changed, from one picture to another (from top left to bottom right), parameters are not restated.}
\label{keen5}
\end{figure}

The Figure \ref{keen5} shows the time evolution of the absolute value of the first Fourier modes of $\rho$.
We see that single precision can modify the results on the long time (top left). The GPU code is validated in double precision (top right).
We clearly see the benefit of the $\delta f$ method in the GPU single precision (middle left), where it has no effect in the double precision case (middle right).
Further plots are given with $N=4096$ (bottom left and right). With smaller time steps, some small oscillations appear with single precision GPU code
(bottom right). In all the plots, we see no difference at the beginning; differences appear in the long run as it was the case for the plots of the distribution function.

\subsection{Performance results}
\label{perf}

\paragraph{\bf Characteristics}
We have tested the code on different computers with the following characteristics:

\begin{itemize}
\item GPU

\begin{itemize}

\item (1) = irma-gpu1 : NVIDIA GTX 470 1280 Mo 

\item (2) = hpc : GPU NVIDIA TESLA C2070

\item (3) = MacBook : NVIDIA GeForce 9400M

\end{itemize}

\item CPU 

\begin{itemize}

\item (4) = MacBook : Intel Core 2 Duo 2.4 GHz

\item (5) = irma-hpc2: Six-Core AMD Opteron(tm) Processor 8439 SE
   
\item (6) = irma-gpu1: Intel Pentium Dual Core 2.2 Ghz 2Gb RAM

\item (7) = MacBook : Intel Core i5 2.4 GHz

\end{itemize}
\end{itemize}
\noindent We measure in the GPU codes the proportion of FFT which consists in: transform $1D$ real data to complex,
computing the FFT, making the complex multiplication, computing the inverse FFT, transforming to real data (together with addition of $\delta f$ modification,
 if we use the $\delta f$ method). We add a diagnostic for having the proportion of time in the \texttt{cufftExec} routine; we note that this extra diagnostic can modify a little the time measures
 (when this is the case; new measures are given in brackets, see on Table \ref{table1}).

\noindent When the number of cells grows, the proportion of FFT time in total time grows, as shown on Table \ref{table1}
(KEEN wave test case with $\delta f$ modification) or Table \ref{table2} (KEEN wave test case without $\delta f$ modification).
Note  that the initialisation time and the 2d-diagnostic time are not included in total time.

The results with a CPU code (vlaso) without OpenMP are given on Table \ref{table3}, top; in that code, the Landau test case is run with
$\Delta t/2$ advection in $x$, followed by $\Delta t$ advection in $v$ and $\Delta t/2$ advection in $x$ and the last advection in $x$ of iteration $n$
is merged with the first advection in $x$ of iteration $n+1$.

The results with Selalib (Table \ref{table3}, bottom) \cite{selalib} are obtained with OpenMP. We use 2 threads for (4), 24 threads for (5), 2 threads for (6) and 4 threads for (7). 

In order to compare the performances, we introduce the number MA which represents the number of millions of point advections made per second : 
$MA= \frac{N_{step} \times N_{adv} \times N^2}{10^6 \times \textnormal{Total time}}$
and the number of operations per second (in GigaFLOPS) given by :
\begin{eqnarray*}
GF&= &
\frac{N_{step} \times N_{adv} \times (2N \times 5N\log(N)+6N^2)}{10^9\times \textnormal{Total time}} \qquad \textnormal{with complex data (GPU)}\\
GF&=& 
\frac{N_{step} \times N_{adv} \times (N \times 5N\log(N)+3N^2)}{10^9\times \textnormal{Total time}} \qquad \textnormal{with real data (CPU)}
\end{eqnarray*}
where $N_{step}$ refers to the number of time steps and $N_{adv}$ represents the number of advections made in each time step ($N_{adv}=3$ in GPU and Selalib codes; $N_{adv}=2$ in vlaso code). In each advection, we compute $N$ times (GPU in complex data) or $N/2$ times (CPU in real data)   :
\begin{itemize}
\item A forward FFT and backward FFT with approximately $5N\log(N)$ operations for each FFT computation
\item A complex multiplication that requires $6N$ operations.
\end{itemize}

\renewcommand{\arraystretch}{0.6}

The comparison between the Table \ref{table1} and the Table \ref{table2} shows that the cost of the method $\delta f$ is not too important but not negligible. This cost could be optimized.
We clearly benefit of the huge acceleration of the FFT routines in GPU and we thus gain a lot to use this approach. 
Most of the work is on the FFT, which is optimized for CUDA in the cufft library, and is transparent for the user.
Note that we are limited here to $N=4096$ in single precision and $N=2048$ in double precision; also we use complex Fourier transform; optimized real transforms may permit to go even faster.
The merge of two velocity time steps can also easily improve the speed. Higher order time splitting may be also used. 
Also, a better comparison with CPU parallelized codes can be envisaged (here, we used a basic OpenMP implementation which only scales for $2$ processors).
We can also hope to go to higher grids, since cufft should allow grid sizes of $128$ millions elements in double precision and $64$ millions in single precision
(here we use $2^{24}\simeq 16.78\cdot 10^{6}$ elements in single precision and $2^{22}\simeq 4.2\cdot 10^{6}$ elements in double precision; so we should be able to run with
$N=8192$ in single precision and $N=4096$ in double precision).
 Higher complexity problems (as $4D$ simulations) will probably need multi-gpu which is another story, see \cite{dannert} for such a work.

\section{Conclusion}
We have shown that this approach works. Most of the load is carried by the FFT routine, which is optimized for CUDA in the cufft library, leading to huge speed-ups and is 
invisible to the user. Thus, the overhead of implementation time which can be quite significant in other contexts is here reduced, since we use largely built-in
routines which are already optimized. The use of single precision is made harmless thanks to a $\delta f$ method. We however are not able to get as precise results as in the case
of double precision. The test cases we chose are quite sensitive to single precision round off errors. We point out also that the electric field has to satisfy a zero mean condition
with enough accuracy on a discrete grid. For the moment, we are limited to same sizes in $x$ and $v$ (needed here for the transposition step) and to $N=2048$ in double precision
($N=4096$ in single precision). We hope to implement a four dimensional (2x, 2v) version of this code, next, including weak collisions. 

\newpage
{\small 
\begin{table}[h]
\begin{center}
\begin{tabular}{|c|c||c|c|c||c|c|c||}
\cline{3-8}
\multicolumn{2}{c||}{ } & \multicolumn{3}{|c||}{ Single precision }& \multicolumn{3}{|c||}{ Double precision } \\
\cline{2-8}
\multicolumn{1}{c|}{ } & $N_x$ & Time (ms)  {(\bf speedup)} & MA & FFT (cufftExec) &  Time (ms) {(\bf speedup)} & MA & FFT (cufftExec) \\
\cline{2-8}
\hline
\hline
&256 & 703 {(\bf 2.8-8.5)}& 279.6 & 0.635 (0.36) & 1304 {\bf(1.5-4.6) }& 150.7 & 0.767 (0.61) \\
\cline{2-8}
&512 & 1878 {(\bf 4.3-17)} & 418.7 & 0.759 (0.46) & 3516 {(\bf 2.3-8.8)}& 223.6 & 0.839 (0.67) \\
\cline{2-8}
(1) &1024 & 6229 {(\bf 9.6-20)} & 505.0 & 0.841 (0.51) & 11670 {(\bf 5.1-11)}& 269.5 & 0.889 (0.71) \\
\cline{2-8}
&2048 & 21908 {(\bf 13-27)} & 574.3 & 0.861 (0.50) & 49925 {(\bf 5.7-12)} & 252.0 & 0.916 (0.75) \\
\cline{2-8}
&4096 & 90093 {(\bf 15-52)} & 558.6 & 0.888 (0.54) & -& - & - \\
\hline
\hline
&256 & 1096 [1378]  {(\bf 1.8-5.5)} & 179.3 & 0.471 [0.59 (0.37)] & 1653 {(\bf 1.2-3.6)} & 118.9 & 0.637 (0.5) \\
\cline{2-8}
&512 & 2125 [2550] {(\bf 3.8-15)}& 370.0 & 0.654 [0.69 (0.48)] & 3896 {(\bf 2.1-8.0)} & 201.8 & 0.777 (0.66) \\
\cline{2-8}
(2) &1024 & 5684 [6001] {(\bf 11-22)} & 553.4 & 0.775 [0.79 (0.59)] & 12127 {(\bf 4.9-10)} & 259.3 & 0.866 (0.76) \\
\cline{2-8}
&2048 & 19871 [20284] {(\bf 14-29)} & 633.2 & 0.825 (0.62) & 45753 {(\bf 6.3-13)} & 275.0 & 0.897 (0.80) \\
\cline{2-8}
&4096 & 81943  {(\bf 17-57)} & 614.2 & 0.859 (0.66) & - & - & - \\
\hline
\hline
&256 & 5783 {(\bf 0.3-1.0)} & 33.9 & 0.773 (0.65) & - & - & -\\
\cline{2-8}
(3) &512 & 19936 {(\bf 0.4-1.6)} & 39.4 & 0.780 (0.66) & - & - & -\\
\cline{2-8}
&1024 & 87685 {(\bf 0.68-1.4)} & 35.8 & 0.813 (0.71) & - & - & -\\
\hline
\end{tabular}
\end{center}
\caption{Performance results for GPU, nbstep=1000, LAG17, KEEN wave test case with $\delta f$ modification:
total time, speedup, MA, proportion FFT/total time (and cufftExec/total time).}
\label{table1}
\end{table}

\begin{table}[h]
\begin{center}
\begin{tabular}{|c|c||c|c|c|c||c|c|c|c||}
\cline{3-10}
\multicolumn{2}{c||}{ } & \multicolumn{4}{|c||}{ Single precision }& \multicolumn{4}{|c||}{ Double precision } \\
\cline{2-10}
\multicolumn{1}{c|}{ } & $N_x$ & Time (ms) {\bf speedup} & MA & GF & FFT & Time (ms)  {\bf speedup} & MA & GF & FFT\\
\cline{2-10}
\hline
\hline
&256 & 570 {(\bf 3.5-11)} & 344.9 & 29.6 & 0.573 & 1183 {(\bf 1.7-5.1)}& 166.1 & 14.2 & 0.754 \\
\cline{2-10}
&512 & 1421 {(\bf 5.6-22)} & 553.4 & 53.1 & 0.702 & 3121 {(\bf 2.6-10)} & 251.9 & 24.1 & 0.826 \\
\cline{2-10}
(1) &1024 & 4516 {(\bf 13-28)} & 696.5 & 73.8 & 0.787 & 10221 {(\bf 5.9-12)} & 307.7 & 32.6 & 0.876 \\
\cline{2-10}
&2048 & 15189 {(\bf 19-38)} & 828.4 & 96.0 & 0.802 & 44244 {(\bf 6.5-13)} & 284.3 & 32.9 & 0.906 \\
\cline{2-10}
&4096 & 63310 {(\bf 22-73)} & 795.0 & 100.1 & 0.842 & - & - & - & -\\
\hline
\hline
&256 & 1000 {(\bf 2.0-6.0)}& 196.6 & 16.9 & 0.520 & 1569 {(\bf 1.3-3.8)}& 125.3 & 10.7 & 0.657 \\
\cline{2-10}
&512 & 2000 {(\bf 4.0-15)} & 393.2 & 37.7 & 0.635 & 3750 {(\bf 2.1-8.3)}& 209.7 & 20.1 & 0.782 \\
\cline{2-10}
(2) &1024 & 5067 {(\bf 12-25)} & 620.8 & 65.8 & 0.762 & 11749 {(\bf 5.1-11)} & 267.7 & 28.3 & 0.865 \\
\cline{2-10}
&2048 & 17692 {(\bf 16-33)} & 711.2 & 82.5 & 0.805 & 44446 {(\bf 6.5-13)} & 283.1 & 32.8 & 0.895 \\
\cline{2-10}
&4096 & 73488 {(\bf 19-63)} & 684.8 & 86.2 & 0.843 & - & - & - & -\\
\hline
\hline
&256 & 5513 {(\bf 0.36-1.1)}& 35.6 & 3.0 & 0.763 & - & - & - & - \\
\cline{2-10}
(3) &512 & 18805 {(\bf 0.43-1.6)} & 41.8 & 4.0 & 0.769 & - & - & - & - \\
\cline{2-10}
&1024 & 83312 {(\bf 0.72-1.5)} & 37.7 & 4.0 & 0.804 & - & - & - & -\\
\hline
\end{tabular}
\end{center}
\caption{Performance results for GPU, nbstep=1000, LAG17, KEEN wave test case without $\delta f$ modification:
total time, speedup, MA, GFlops and proportion FFT/total time.
}
\label{table2}
\end{table}
\begin{table}[h]
\begin{center}
\begin{tabular}{|c||c|c|c||c|c|c||c|c|c||c|c|c||}
\cline{2-13}
\multicolumn{1}{c||}{ } & \multicolumn{3}{|c||}{ (4) }& \multicolumn{3}{|c||}{ (5) } & \multicolumn{3}{|c||}{ (6) } & \multicolumn{3}{|c||}{ (7) } \\
\hline
 $N_x$ & Total time & MA & GF & Total time & MA & GF & Total time & MA & GF & Total time & MA & GF \\
\hline
\hline
256 & 4s & 27.4 & 1.1 & 4s & 28.8 & 1.2 & 6s & 21.4 & 0.9 & 3s & 38.8 & 1.6\\
\hline
512 & 27s & 19.2 & 0.9 & 18s & 28.8 & 1.3 & 31s & 16.5 & 0.7 & 15s & 34.7 & 1.6\\
\hline
1024 & 1min52s & 18.7 & 0.9 & 2min4s & 16.8 & 0.8 & 2min7s & 16.4 & 0.8 & 1min18s & 26.7 & 1.4\\
\hline
2048 & 8min16s & 16.9 & 0.9 & 9min31s & 14.6 & 0.8 & 9min42s & 14.4 & 0.8 & 5min36s & 24.9 & 1.4\\
\hline
4096 & 41min05s & 13.6 & 0.8 & 48min16s & 11.5 & 0.7 & 52min20s & 10.6 & 0.6 & 28min28s & 19.6 & 1.2\\
\hline
\hline
256 & 3s & 58.0 & 2.4 & 4s & 43.9 & 1.8 & 3s & 54.3 & 2.3 & 2s & 72.6 & 3.1\\
\hline
512 & 19s & 39.6 & 1.9 & 8s & 90.6 & 4.3 & 22s & 35.0 & 1.6 & 13s & 58.7 & 2.8\\
\hline
1024 & 1min25s & 36.8 & 1.9 & 1min21s & 38.5 & 2.0 & 1min35s & 32.9 & 1.7 & 1min0s & 52.1 & 2.7\\
\hline
2048 & 6min41s & 31.3 & 1.8 & 7min46s & 27.0 & 1.5 & 8min47s & 28.3 & 1.6 & 4min47s & 43.7 & 2.5\\
\hline
4096 & 34min39s & 24.2 & 1.5 & 25min33s & 32.8 & 2.0 & 77min31s & 10.8 & 0.6 & 23min09s & 36.2 & 2.2\\
\hline
\end{tabular}
\end{center}
\caption{ Performance results for CPU vlaso code,  nbstep=1000, LAG 17, Landau test case (top):
total time, MA and GFlops.
Performance results for CPU Selalib code,  nbstep=1000, LAG 17, KEEN test case without $\delta f$ modification (bottom):
total time, MA and GFlops.
}
\label{table3}
\end{table}
}


\newpage

\begin{thebibliography}{999}
%
\bibitem{bedros}
{\sc B. Afeyan, K. Won, V. Savchenko, T. Johnston, A. Ghizzo, and P. Bertrand.}
\textit{ Kinetic
Electrostatic Electron Nonlinear (KEEN) Waves and their Interactions Driven by the
Ponderomotive Force of Crossing Laser Beams.}, Proc. IFSA 2003, 213, 2003, and
arXiv:1210.8105, \url{http://arxiv.org/abs/1210.8105}.
 	
\bibitem{arber}
{\sc T. D. Arber, R. G. Vann,} 
\textit{A critical comparison of Eulerian-grid-based Vlasov solvers}, 
JCP, {\bf 180} (2002),  pp. 339-357.

\bibitem{BeMe}
{\sc N. Besse, M. Mehrenberger,}
\textit{Convergence of classes of high-order semi-lagrangian schemes for the Vlasov-Poisson system,}
Mathematics of Computation, 77, 93--123 (2008).


\bibitem{birdsall}
{\sc C. K. Birdsall, A. B. Langdon}, 
\textit{Plasma Physics via Computer Simulation}, Adam Hilger, 1991.

\bibitem{picgpu}
{\sc K. J. Bowers, B. J. Albright, B. Bergen, L. Yin, K. J. Barker,  D. J. Kerbyson},  
\textit{$0.374$ pflop/s trillion-particle kinetic modeling of laser plasma interaction on roadrunner},  
Proc. of Supercomputing. IEEE Press, 2008.

\bibitem{boris}
{\sc J. P. Boris, D. L. Book,} 
\textit{Flux-corrected transport. I: SHASTA, a fluid transport algorithm that works}, 
J. Comput. Phys. {\bf 11} (1973), pp. 38-69. 
%
\bibitem{ChaDeMe2012}
{\sc F. Charles, B. Despr\'es, M. Mehrenberger,}
\textit{Enhanced convergence estimates for semi-lagrangian schemes Application to the Vlasov-Poisson equation, }
accepted in SINUM, and inria-00629081, version 1, October 2011, \url{http://hal.inria.fr/inria-00629081/}.


\bibitem{cgm2012}
{\sc Y. Cheng, I. M. Gamba, P. J. Morrison,}
\textit{Study of conservation and recurrence of Runge-Kutta
discontinuous Galerkin schemes for Vlasov-Poisson
systems},arXiv:1209.6413v2, 17 Dec 2012, \url{http://arxiv.org/abs/1209.6413}.

\bibitem{cheng}
{\sc C. Z. Cheng, G. Knorr}, 
\textit{The integration of the Vlasov equation in configuration space}, 
J. Comput. Phys. {\bf 22} (1976),  pp. 330-3351. 


\bibitem{anais}
{\sc A. Crestetto, P. Helluy},
\textit{Resolution of the Vlasov-Maxwell system by PIC Discontinuous Galerkin method on GPU with OpenCL},
\url{http://hal.archives-ouvertes.fr/hal-00731021}

\bibitem{cfm}
{\sc N. Crouseilles, E. Faou, M. Mehrenberger}, 
\textit{High order Runge-Kutta-Nystr\"om splitting methods for the Vlasov-Poisson equation}, 
inria-00633934, version 1, \url{http://hal.inria.fr/IRMA/inria-00633934}.


\bibitem{crous}
{\sc N. Crouseilles, M. Mehrenberger, E. Sonnendr\"ucker}, 
\textit{Conservative semi-Lagrangian schemes for Vlasov equations}, 
J. Comput. Phys. {\bf 229} (2010), pp. 1927-1953. 


\bibitem{dannert}
{\sc T. Dannert}, 
\textit{GENE on Accelerators},  4th Summer school on numerical modeling for fusion, 8-12 October 2012, IPP, Garching near Munich, Germany, \url{http://www.ipp.mpg.de/ippcms/eng/for/veranstaltungen/konferenzen/su\_school/}




%
\bibitem{fijalkow}
{\sc E. Fijalkow,} 
\textit{A numerical solution to the Vlasov equation}, 
Comput. Phys. Commun. {\bf 116} (1999), pp. 329Ð 335.
%
\bibitem{filbet}
{\sc F. Filbet, E. Sonnendr\"{u}cker, P. Bertrand}, 
\textit{Conservative numerical schemes for the Vlasov equation},
J. Comput. Phys.  {\bf 172}  (2001), pp. 166-187. 
%
\bibitem{filbet2}
{\sc F. Filbet, E. Sonnendr\"{u}cker}, 
\textit{Comparison of  Eulerian Vlasov solvers},
Comput. Phys. Comm.  {\bf 151} (2003), pp. 247-266.
%

\bibitem{filbet_landau}
{\sc F. Filbet}
\textit{Numerical simulations avalaible online at}
\url{http://math.univ-lyon1.fr/\~filbet/publication.html}
\bibitem{gray}
{\sc R.M. Gray}, 
\textit{Toeplitz and circulant matrices: a review}, 
Now Publishers Inc, Boston-Delft  (2005)

\bibitem{guclu}
{\sc Y. Guclu, W. N. G. Hitchon, Szu-Yi Chen,}
\textit{High order semi-lagrangian methods for the kinetic description of plasmas,} 
Plasma Science (ICOPS), 2012 Abstracts IEEE, vol., no., pp.5A-5, 8-13 July 2012, 
doi: 10.1109/PLASMA.2012.6383976.


\bibitem{hatzky}
{\sc R. Hatzky}, 
\textit{Global electromagnetic gyrokinetic particle-in-cell simulation},  4th Summer school on numerical modelling for fusion, 8-12 October 2012, IPP, Garching near Munich, Germany, \url{http://www.ipp.mpg.de/ippcms/eng/for/veranstaltungen/konferenzen/su\_school/}.

\bibitem{krall}
{\sc N.A. Krall, A.W. Trivelpiece},
\textit{Principles of Plasma Physics}, McGrawÐHill, New York (1973).

\bibitem{latu-gpu}
{\sc G. Latu},  
\textit{Fine-grained parallelization of Vlasov-Poisson application on GPU}, 
Euro-Par 2010, Parallel Processing Workshops, Springer (New York, 2011).
%

\bibitem{villani}
{\sc C. Mouhot, C. Villani},
\textit{On Landau damping}, Acta Mathematica, volume 207, number 1, pages 29-201, september 2011, arXiv:0904.2760,
\url{http://arxiv.org/abs/0904.2760}.

\bibitem{qiushu}
{\sc J.M. Qiu, C. W. Shu}, 
\textit{Conservative semi-Lagrangian finite difference WENO formulations with applications to the Vlasov equation},
Comm.  Comput. Phys. {\bf 10} (2011), pp 979-1000.


\bibitem{filho}
{\sc T. M. Rocha Filho}, 
\textit{Solving the Vlasov equation for one-dimensional models with long range interactions on a GPU}, 
\url{http://arxiv.org/abs/1206.3229}.

\bibitem{selalib}
\textit{Selalib, a semi-Lagrangian library,} \url{http://selalib.gforge.inria.fr/}

\bibitem{shoucri}
{\sc M. Shoucri}, 
\textit{Nonlinear evolution of the bump-on-tail instability}, 
Phys. Fluids {\bf 22} (1979), pp. 2038-2039. 


\bibitem{keen_sonnen}
{\sc E. Sonnendr\"ucker , N. Crouseilles , B. Afeyan}, 
\textit{BP8.00057: High Order Vlasov Solvers for the Simulation of KEEN Wave Including the L-B and F-P Collision Models},
54th Annual Meeting of the APS Division of Plasma Physics Volume 57, Number 12, MondayÐFriday, October 29--November 2 2012; Providence, Rhode Island,
\url{http://meeting.aps.org/Meeting/DPP12/SessionIndex2/?SessionEventID=181483}.


\bibitem{sonnen}
{\sc E. Sonnendr\"ucker}, 
\textit{Approximation num\'erique des \'equations de Vlasov-Maxwell},
Master lectures, \url{http://www-irma.u-strasbg.fr/~sonnen/polyM2VM2010.pdf}.

\bibitem{pic-gpu}
{\sc G. Stantchev, W. Dorland, N. Gumerov},  
\textit{Fast parallel
particle-to-grid interpolation for plasma PIC simulations on the GPU},  
J. Parallel Distrib. Comput., {\bf 68}(10), pp. 1339-1349, (2008).

\bibitem{zhou} 
{\sc T. Zhou, Y. Guo, C.W. Shu},
\textit{Numerical study on Landau damping}, Physica D 157 (2001), 322--333.

\end{thebibliography}
\end{document}